\documentclass[12pt]{article}
\usepackage{amsmath,float,amssymb,amsthm,amsxtra,overpic,bbm,bm,epsfig}
\def\thefootnote{\fnsymbol{footnote}}

\begin{document}

\vspace{0.2cm}

\begin{center}
{\large\bf Analytical approximations for matter effects
on CP violation in \\ the accelerator-based neutrino oscillations
with $E \lesssim 1$ GeV}
\end{center}

\vspace{0.1cm}

\begin{center}
{\bf Zhi-zhong Xing}$^{1, 2, 3}$ \footnote{E-mail: xingzz@ihep.ac.cn}
~and~ {\bf Jing-yu Zhu}$^{1}$
\footnote{E-mail: zhujingyu@ihep.ac.cn} \\
{\small $^1$Institute of High Energy Physics, Chinese Academy of
Sciences, Beijing 100049, China \\
$^2$School of Physical Sciences, University of Chinese Academy of
Sciences, Beijing 100049, China \\
$^3$Center for High Energy Physics, Peking University, Beijing
100080, China }
\end{center}

\vspace{1.5cm}

\begin{abstract}
Given an accelerator-based neutrino experiment with the beam energy
$E\lesssim 1$ GeV, we expand the probabilities of $\nu^{}_\mu \to
\nu^{}_e$ and $\overline{\nu}^{}_\mu \to \overline{\nu}^{}_e$
oscillations in matter in terms of two small quantities
$\Delta^{}_{21}/\Delta^{}_{31}$ and $A/\Delta^{}_{31}$, where
$\Delta^{}_{21} \equiv m^2_2 - m^2_1$ and $\Delta^{}_{31} \equiv
m^2_3 - m^2_1$ are the neutrino mass-squared differences, and $A$
measures the strength of terrestrial matter effects. Our analytical
approximations are numerically more accurate than those made by
Freund in this energy region, and thus they are particularly
applicable for the study of leptonic CP violation in the low-energy
MOMENT, ESS$\nu$SM and T2K oscillation experiments. As a by-product,
the new analytical approximations help us to easily understand why
the matter-corrected Jarlskog parameter $\widetilde{\cal J}$ peaks
at the resonance energy $E^{}_* \simeq 0.14$ GeV (or $0.12$ GeV) for
the normal (or inverted) neutrino mass hierarchy, and how the three
Dirac unitarity triangles are deformed due to the terrestrial matter
contamination. We also affirm that a medium-baseline neutrino
oscillation experiment with the beam energy $E$ lying in the $E^{}_*
\lesssim E \lesssim 2 E^{}_*$ range is capable of exploring leptonic
CP violation with little matter-induced suppression.
\end{abstract}

\begin{flushleft}
\hspace{0.8cm} PACS number(s): 14.60.Pq, 13.10.+q, 25.30.Pt \\
\hspace{0.8cm} Keywords: CP violation, matter effects, unitarity
triangles, neutrino oscillations
\end{flushleft}

\def\thefootnote{\arabic{footnote}}
\setcounter{footnote}{0}

\newpage

\section{Introduction}

In the past two decades we have witnessed a booming period in
neutrino physics thanks to a number of indisputable observations of
atmospheric, solar, reactor and accelerator neutrino oscillations
\cite{PDG}, and thus achieved a smoking gun for the incompleteness
of the standard model (SM) in particle physics --- the neutrinos
actually have finite rest masses and the lepton flavors are
significantly mixed, motivating us to explore the other unknowns of
massive neutrinos beyond the SM and search for their possible
consequences in nuclear physics, particle astrophysics and
cosmology.

In the standard three-flavor scheme there are six neutrino
oscillation parameters: two independent neutrino mass-squared
differences (e.g., $\Delta^{}_{21} \equiv m^2_2 - m^2_1$ and
$\Delta^{}_{31} \equiv m^2_3 - m^2_1$), three lepton flavor mixing
angles (i.e., $\theta^{}_{12}$, $\theta^{}_{13}$ and
$\theta^{}_{23}$) and one CP-violating phase (i.e., $\delta$). Among
them, the sign of $\Delta^{}_{31}$ and the size of $\delta$ remain
unknown \cite{Fogli,Valle,GG}. But some preliminary hints for
$\delta \sim 3\pi/2$ and $\Delta^{}_{31} > 0$ have recently been
seen by combining the T2K \cite{T2K,T2K2} and NO$\nu$A \cite{NOVA}
data on $\nu^{}_\mu \to \nu^{}_e$ oscillations with the Daya Bay
(reactor $\overline{\nu}^{}_e \to \overline{\nu}^{}_e$ oscillation
\cite{DYB,DYB2}) and Super-Kamiokande (atmospheric $\nu^{}_\mu \to
\nu^{}_\mu$ oscillation \cite{SK}) data \cite{Lisi}. Provided
$\delta$ is really around $3\pi/2$ or takes a nontrivial value far
away from $0$ and $\pi$, then remarkable CP- and T-violating effects
will emerge in some upcoming long-baseline neutrino oscillation
experiments.

Among a number of ongoing and proposed accelerator-based experiments
which aim to probe or constrain CP violation in neutrino
oscillations \cite{Wang}, those with the beam energy $E \lesssim 1$
GeV (e.g., T2K \cite{T2K}, MOMENT \cite{MOMENT} and ESS$\nu$SM
\cite{ESS}) are expected to involve much smaller terrestrial matter
effects. To understand the salient features of the matter-corrected
$\nu^{}_\mu \to \nu^{}_e$ and $\overline{\nu}^{}_\mu \to
\overline{\nu}^{}_e$ oscillations in this energy region, it is
important to expand their probabilities in terms of two small
expansion parameters $\alpha \equiv \Delta^{}_{21}/\Delta^{}_{31}$
and $\beta \equiv A/\Delta^{}_{31}$, where $A \equiv 2 \sqrt{2} \
G^{}_{\rm F} N^{}_e E$ with $G^{}_{\rm F}$ being the Fermi constant
and $N^{}_e$ being the background density of electrons. But the
previous analytical approximations in this connection, such as the
popular one developed by Freund \cite{Freund}, are usually subject
to $E \gtrsim 0.5$ GeV and will become invalid when $E$ approaches
vanishing
\footnote{Xu has noticed that the approximate formulas obtained by
Freund \cite{Freund} are still valid even near the solar neutrino
resonance in matter (i.e., $A \simeq \Delta^{}_{21} \cos
2\theta^{}_{12}$) \cite{Xu}, but we are going to show that they will
become problematic in the $E \lesssim 0.4$ GeV region and definitely
turn to be invalid in the $E \lesssim 0.1$ GeV region.}.
The reason is simply that mainly the long-baseline neutrino
oscillation experiments with $E \gtrsim 1$ GeV were considered
in those works.

Hence our present work is well motivated to offer the hitherto most
systematic and useful analytical approximations for terrestrial
matter effects on CP violation in the medium-baseline neutrino
oscillation experiments with the beam energy $E \lesssim 1$ GeV.

The strength of CP and T violation in neutrino oscillations is
measured by a universal and rephasing-invariant quantity of the
$3\times 3$ Pontecorvo-Maki-Nakagawa-Sakata (PMNS) lepton flavor
mixing matrix $U$ \cite{PMNS,PMNS2,PMNS3}, the so-called Jarlskog
parameter $\cal J$ \cite{J} defined via
\begin{eqnarray}
{\rm Im}\left(U^{}_{\alpha i} U^{}_{\beta j} U^*_{\alpha j}
U^*_{\beta i}\right) = {\cal J} \sum_\gamma
\epsilon^{}_{\alpha\beta\gamma} \sum_k \epsilon^{}_{ijk} \; ,
\end{eqnarray}
where the Greek and Latin subscripts run over $(e, \mu, \tau)$ and
$(1,2,3)$, respectively. When a neutrino beam travels through a
medium, it can see two kinds of refractive indices because of its
interactions with the constituents of the medium (i.e., electrons,
protons and neutrons) via the weak neutral current (NC) and charged
current (CC) \cite{MSW,MSW2}. All the three neutrino flavors share a
common ``matter" phase due to the refractive index arising from the
NC forward scattering, but the electron neutrinos develop an extra
``matter" phase owing to the CC forward scattering. The latter is
nontrivial, and hence it is likely to change the neutrino
oscillation behavior. In this case one may define the
matter-corrected neutrino masses $\widetilde{m}^{}_i$ and the
corresponding PMNS matrix $\widetilde{U}$, so as to express the
probabilities of neutrino oscillations in matter in the same way as
those in vacuum. For example, the T-violating asymmetry between the
probabilities of $\nu^{}_\mu \to \nu^{}_e$ and $\nu^{}_e \to
\nu^{}_\mu$ oscillations in matter is given by \cite{Petcov,Ohlsson}
\footnote{Since an ordinary medium (e.g., the Earth) only consists
of electrons, protons and neutrons instead of both these particles
and their antiparticles, the matter background is not symmetric
under the CP transformation. Hence the expression of the
CP-violating asymmetry between $P(\nu^{}_\mu \to \nu^{}_e)$ and
$P(\overline{\nu}^{}_\mu \to \overline{\nu}^{}_e)$ is not so simple
as that of $\widetilde{\cal A}^{}_{\rm T}$ in Eq. (2), as one can
clearly see in section 4.}
\begin{eqnarray}
\widetilde{\cal A}^{}_{\rm T} = -16 \widetilde{\cal J}
\sin\frac{\widetilde{\Delta}^{}_{21} L}{4E}
\sin\frac{\widetilde{\Delta}^{}_{31} L}{4E}
\sin\frac{\widetilde{\Delta}^{}_{32} L}{4E} \; ,
\end{eqnarray}
in which $E$ denotes the neutrino beam energy, $L$ is the distance
between a neutrino source and the detector, $\widetilde{\cal J}$ and
$\widetilde{\Delta}^{}_{ij}$ are the matter-corrected counterparts
of ${\cal J}$ and $\Delta^{}_{ij}$ (for $ij = 21, 31, 32$),
respectively. It is known that $\widetilde{\cal J}
\widetilde{\Delta}^{}_{21} \widetilde{\Delta}^{}_{31}
\widetilde{\Delta}^{}_{32} = {\cal J} \Delta^{}_{21} \Delta^{}_{31}
\Delta^{}_{32}$ exactly holds for a constant matter profile
\cite{Naumov,Naumov2,Naumov3}. But a more transparent relationship
between $\widetilde{\cal J}$ and $\cal J$, which can directly tell
us why or how CP violation in matter is enhanced or suppressed as
compared with that in vacuum, has been lacking. It should be noted
that $\cal J$ (or $\widetilde{\cal J}$) is in principle a measurable
quantity, but in practice it is not directly observable since it is
always correlated with the oscillation terms as shown in Eq. (2).

However, a careful study of the ratio $\widetilde{\cal J}/{\cal J}$
changing with the neutrino (or antineutrino) beam energy $E$ is not
only conceptually interesting but also practically indispensable for
expanding the matter-corrected oscillation probabilities
$\widetilde{P}(\nu^{}_\mu \to \nu^{}_e)$ and
$\widetilde{P}(\overline{\nu}^{}_\mu \to \overline{\nu}^{}_e)$ in
terms of the afore-defined small parameters $\alpha$ and $\beta$ in
the $E \lesssim 1$ GeV region. So we plan to organize the remaining
parts of this paper in an easy-to-follow and step-by-step way:
starting from the analytical approximation of $\widetilde{\cal
J}/{\cal J}$, passing through those of $|\widetilde{U}^{}_{e i}
\widetilde{U}^*_{\mu i}|$, $|\widetilde{U}^{}_{\mu i}
\widetilde{U}^*_{\tau i}|$ and $|\widetilde{U}^{}_{\tau i}
\widetilde{U}^*_{e i}|$ (for $i=1,2,3$), and ending with those of
$\widetilde{P}(\nu^{}_\mu \to \nu^{}_e)$ and
$\widetilde{P}(\overline{\nu}^{}_\mu \to \overline{\nu}^{}_e)$.

In section 2 we aim to reveal a unique range of the
{\it neutrino} beam energy $E$ in which the size of the effective
Jarlskog invariant $\widetilde{\cal J}$ can be enhanced as compared
with its fundamental counterpart $\cal J$. We find that
$\widetilde{\cal J}/{\cal J} \gtrsim 1$ will hold if $E$ is below
the upper limit $E^{}_0 \simeq \Delta^{}_{21} \cos 2\theta^{}_{12}
/(\sqrt{2} \ G^{}_{\rm F} N^{}_e) \lesssim 0.3$ GeV in a realistic
oscillation experiment. In particular,
we find that $\widetilde{\cal J}/{\cal J}$ peaks at the resonance
energy
\begin{eqnarray}
E^{}_* \simeq \frac{\Delta^{}_{21}}{2 \sqrt{2} \ G^{}_{\rm F}
N^{}_e} \left[ \cos 2\theta^{}_{12} \left(1 + \sin^2
\theta^{}_{13}\right) + \alpha \sin^2 2\theta^{}_{12} \right] \; ,
\end{eqnarray}
which is about $0.14$ GeV (or $0.12$ GeV) for $\Delta^{}_{31} > 0$
(or $\Delta^{}_{31} <0$), corresponding to the normal (or inverted)
neutrino mass ordering. Accordingly, we arrive at the maximum value
\begin{eqnarray}
\frac{\widetilde{\cal J}^{}_*}{\cal J} \simeq
\frac{1}{\sin 2\theta^{}_{12}} \left[1 + \alpha \cos 2 \theta_{12}
^{} \left( 1 + \sin^2 \theta_{13}^{} \right) + {\rm smaller ~
terms}  \right] \; ,
\end{eqnarray}
which is roughly $110\%$ (or $107\%$) for $\Delta^{}_{31} > 0$ (or
$\Delta^{}_{31} <0$). As for an {\it antineutrino} beam,
$\widetilde{\cal J}/{\cal J}$ decreases monotonically in the $E
\lesssim 1$ GeV region and thus does not undergo any resonances. In
this sense one may draw the conclusion that a medium-baseline
neutrino oscillation experiment with $E$ being in the range $E^{}_*
\lesssim E \lesssim 2 E^{}_*$ should be able to explore leptonic CP
violation with little matter-induced suppression
\footnote{Note that Minakata and Nunokawa have discussed a similar
possibility and obtained the leading-order analytical result of
$E^{}_*$ in Ref. \cite{Minakata}. In comparison, our analytical
result in Eq. (3) has a much higher degree of accuracy and thus the
new result in Eq. (4) can explain the sensitivity of
$\widetilde{\cal J}^{}_*/{\cal J}$ to the neutrino mass ordering.}.

In section 3 we concentrate on a geometrical description of leptonic
CP violation in matter and make some analytical approximations for
this intuitive and useful language. Namely, we show how the three
Dirac unitarity triangles (UTs) in the complex plane \cite{FX00}
\footnote{The other three unitarity triangles (defined as
$\triangle^{}_1$, $\triangle^{}_2$ and $\triangle^{}_3$), the
so-called Majorana UTs \cite{Branco,Zhu}, will not be discussed here
because they have nothing to do with leptonic CP and T violation in
normal neutrino-neutrino and antineutrino-antineutrino
oscillations.},
defined through the orthogonality relations
\begin{eqnarray}
\triangle^{}_e : && U^{}_{\mu 1} U^*_{\tau 1} + U^{}_{\mu 2}
U^*_{\tau 2} + U^{}_{\mu 3} U^*_{\tau 3} = 0 \; ,
\nonumber \\
\triangle^{}_\mu : && U^{}_{\tau 1} U^*_{e 1} + U^{}_{\tau 2} U^*_{e
2} + U^{}_{\tau 3} U^*_{e 3} = 0 \; ,
\nonumber \\
\triangle^{}_\tau : && U^{}_{e 1} U^*_{\mu 1} + U^{}_{e 2} U^*_{\mu
2} + U^{}_{e 3} U^*_{\mu 3} = 0 \; , \hspace{1cm}
\end{eqnarray}
are modified (either enlarged or suppressed) by terrestrial matter
effects in a low-energy medium-baseline neutrino oscillation
experiment. We find that the third side of each UT (i.e., $U^{}_{\mu
3} U^*_{\tau 3}$, $U^{}_{\tau 3} U^*_{e 3}$ or $U^{}_{e 3} U^*_{\mu
3}$) is essentially insensitive to the matter-induced corrections
when the neutrino beam energy $E$ is low, but the other two sides
--- both their sizes and orientations --- can get appreciable
corrections. Besides some new and useful analytical results to be
obtained in a reasonably good approximation, a numerical
illustration of the real shapes of the effective Dirac UTs in matter
(denoted as $\widetilde{\triangle}^{}_e$,
$\widetilde{\triangle}^{}_\mu$ and $\widetilde{\triangle}^{}_\tau$)
changing with $E$ will also be presented.

In section 4 we aim to combine our new results about
$\widetilde{\cal J}$ and $\widetilde{\triangle}^{}_\alpha$ (for
$\alpha = e, \mu, \tau$) with the probabilities of neutrino
oscillations in matter. In particular, the effective probabilities
$\widetilde{P}(\nu^{}_\mu \to \nu^{}_e)$ and
$\widetilde{P}(\overline{\nu}^{}_\mu \to \overline{\nu}^{}_e)$ are
expanded in the whole $E \lesssim 1$ GeV region with the help of the
small quantities $\alpha$ and $\beta$. We show that our analytical
approximations are numerically more accurate than those made by
Freund in this energy region, and thus they are particularly
applicable for the study of leptonic CP violation in the low-energy
MOMENT, ESS$\nu$SM and T2K oscillation experiments. We also affirm
that a medium-baseline neutrino oscillation experiment with the beam
energy $E$ lying in the $E^{}_* \lesssim E \lesssim 2 E^{}_*$ range
is capable of exploring leptonic CP violation with little
matter-induced suppression.

\section{The matter-enhanced Jarlskog parameter}

Given the effective neutrino masses $\widetilde{m}^{}_i$ and the
effective lepton flavor mixing matrix $\widetilde{U}$ which have
accommodated the matter-induced corrections to $m^{}_i$ and $U$, the
effective Hamiltonian responsible for the propagation of a neutrino
beam in matter can be written as \cite{MSW,MSW2}
\begin{eqnarray}
\widetilde{\cal H}^{}_{\rm eff} = \frac{1}{2 E} \left[\widetilde{U}
\begin{pmatrix} \widetilde{m}^2_1 & 0 & 0 \cr 0 & \widetilde{m}^2_2
& 0 \cr 0 & 0 & \widetilde{m}^2_3 \cr \end{pmatrix}
\widetilde{U}^\dagger \right] = \frac{1}{2 E} \left[U
\begin{pmatrix} m^2_1 & 0 & 0 \cr 0 & m^2_2 & 0 \cr 0 & 0 & m^2_3
\cr \end{pmatrix} U^\dagger + \begin{pmatrix} A & 0 & 0 \cr 0 & 0 &
0 \cr 0 & 0 & 0 \cr \end{pmatrix} \right] \; ,
\end{eqnarray}
in which $A = 2\sqrt{2} \ G^{}_{\rm F} N^{}_e E$ denotes the
charged-current contribution to the coherent $\nu^{}_e e^-$ forward
scattering in matter. When a constant terrestrial matter profile is
concerned, as in the present work, Eq. (6) allows one to derive the
following relation between the fundamental Jarlskog invariant $\cal
J$ and its matter-corrected counterpart $\widetilde{\cal J}$:
\begin{eqnarray}
\frac{\widetilde{\cal J}}{\cal J} = \left|\frac{\widetilde{U}^{}_{e
1}}{U^{}_{e 1}}\right| \left|\frac{\widetilde{U}^{}_{e 2}}{U^{}_{e
2}}\right| \left|\frac{\widetilde{U}^{}_{e 3}}{U^{}_{e 3}}\right| =
\frac{\Delta^{}_{21} \Delta^{}_{31}
\Delta^{}_{32}}{\widetilde{\Delta}^{}_{21}
\widetilde{\Delta}^{}_{31} \widetilde{\Delta}^{}_{32}} \; ,
\end{eqnarray}
which is a reflection of both the Naumov relation
\cite{Naumov,Naumov2,Naumov3} and the Toshev relation \cite{Toshev}.
The latter means $\sin 2\widetilde{\theta}^{}_{23} \sin
\widetilde{\delta} = \sin 2\theta^{}_{23} \sin\delta$ in the
standard parametrization of $U$ and $\widetilde{U}$. Namely
\footnote{For the sake of simplicity, we have omitted the Majorana
CP-violating phases of massive neutrinos in this parametrization
simply because they have nothing to do with neutrino oscillations
under discussion.},
\begin{eqnarray}
U = \begin{pmatrix} U^{}_{e 1} & U^{}_{e 2} & U^{}_{e 3} \cr
U^{}_{\mu 1} & U^{}_{\mu 2} & U^{}_{\mu 3} \cr
U^{}_{\tau 1} & U^{}_{\tau 2} & U^{}_{\tau 3} \cr \end{pmatrix}
=
\begin{pmatrix} c^{}_{12} c^{}_{13} & s^{}_{12} c^{}_{13} & s^{}_{13}
e^{-{\rm i} \delta} \cr -s^{}_{12} c^{}_{23} - c^{}_{12} s^{}_{13}
s^{}_{23} e^{{\rm i} \delta} & c^{}_{12} c^{}_{23} - s^{}_{12}
s^{}_{13} s^{}_{23} e^{{\rm i} \delta} & c^{}_{13} s^{}_{23} \cr
s^{}_{12} s^{}_{23} - c^{}_{12} s^{}_{13} c^{}_{23} e^{{\rm i}
\delta} & ~ -c^{}_{12} s^{}_{23} - s^{}_{12} s^{}_{13} c^{}_{23}
e^{{\rm i} \delta} ~ & c^{}_{13} c^{}_{23} \cr
\end{pmatrix}
\end{eqnarray}
with $c^{}_{ij} \equiv \cos\theta^{}_{ij}$ and $s^{}_{ij} \equiv
\sin\theta^{}_{ij}$ (for $ij = 12, 13, 23$). The parametrization of
$\widetilde{U}$ is exactly the same as that of $U$ in Eq. (8), and
hence one may obtain ${\cal J} = c^{}_{12} s^{}_{12} c^2_{13}
s^{}_{13} c^{}_{23} s^{}_{23} \sin \delta$ and the same expression
of $\widetilde{\cal J}$ as a function of
$\widetilde{\theta}^{}_{12}$, $\widetilde{\theta}^{}_{13}$,
$\widetilde{\theta}^{}_{23}$ and $\widetilde{\delta}$. Note,
however, that Eq. (7) is actually a parametrization-independent
result. We shall use it to establish an approximate but more
transparent relationship between $\cal J$ and $\widetilde{\cal J}$
later on.

In fact, the exact relations between $\widetilde{m}^2_i$ and $m^2_i$
(for $i=1,2,3$) have been derived by several authors with the help
of Eq. (6) \cite{Barger,Zaglauer,Xing2000}, but only the normal
neutrino mass ordering (i.e., $\Delta^{}_{31} >0$) was assumed in
those works. Here we consider both normal and inverted (i.e.,
$\Delta^{}_{31} <0$) neutrino mass hierarchies. To be explicit, we
have
\begin{eqnarray}
\widetilde{\Delta}^{}_{21} \hspace{-0.15cm} & = &
\hspace{-0.15cm} \frac{2}{3} \sqrt{x^2 - 3 y} \sqrt{3 \left(1 -
z^2\right)} \;\; ,
\nonumber \\
\widetilde{\Delta}^{}_{31} \hspace{-0.15cm} & = &
\hspace{-0.15cm} \frac{1}{3} \sqrt{x^2 - 3 y} \left[3z + \sqrt{3
\left(1 - z^2\right)}\right] \; ,
\nonumber \\
\widetilde{\Delta}^{}_{32} \hspace{-0.15cm} & = &
\hspace{-0.15cm} \frac{1}{3} \sqrt{x^2 - 3 y} \left[3z - \sqrt{3
\left(1 - z^2\right)}\right] \;  \hspace{0.5cm}
\end{eqnarray}
in the $\Delta^{}_{31} >0$ case; or
\begin{eqnarray}
\widetilde{\Delta}^{}_{21} \hspace{-0.15cm} & = &
\hspace{-0.15cm} \frac{1}{3} \sqrt{x^2 - 3 y} \left[3z - \sqrt{3
\left(1 - z^2\right)}\right] \; ,
\nonumber \\
\widetilde{\Delta}^{}_{31} \hspace{-0.15cm} & = &
\hspace{-0.15cm} -\frac{2}{3} \sqrt{x^2 - 3 y} \sqrt{3 \left(1 -
z^2\right)} \;\; ,
\nonumber \\
\widetilde{\Delta}^{}_{32} \hspace{-0.15cm} & = &
\hspace{-0.15cm} -\frac{1}{3} \sqrt{x^2 - 3 y} \left[3z + \sqrt{3
\left(1 - z^2\right)}\right] \;  \hspace{0.3cm}
\end{eqnarray}
in the $\Delta^{}_{31} < 0$ case, where
\begin{eqnarray}
x \hspace{-0.15cm} & = & \hspace{-0.15cm}
\displaystyle \Delta^{}_{31} \left( 1 + \alpha + \beta\right) \; ,
\nonumber \\
y \hspace{-0.15cm} & = & \hspace{-0.15cm}
\displaystyle \Delta^2_{31} \left[ \alpha + \beta
\left(|U^{}_{e 1}|^2 + |U^{}_{e 2}|^2\right) + \alpha\beta \left(1 -
|U^{}_{e 2}|^2\right) \right] \; ,
\nonumber \\
z \hspace{-0.15cm} & = & \hspace{-0.15cm}
\cos\left[ \frac{1}{3} \arccos \frac{\displaystyle 2 x^3 - 9
xy + 27 \Delta^3_{31} \alpha\beta |U^{}_{e 1}|^2}{\displaystyle 2
\sqrt{\displaystyle \left( x^2 - 3 y\right)^3}} \right] \;
\end{eqnarray}
with the definitions $\alpha \equiv \Delta^{}_{21}/\Delta^{}_{31}$
and $\beta \equiv A/\Delta^{}_{31}$. When an {\it antineutrino} beam
is taken into account, the corresponding oscillation behaviors
depend on $\widetilde{U}^*$ and $-A$. In this case the above
formulas remain valid but the replacements $U \to U^*$ and $A \to
-A$ (i.e., $\cal J \to - J$ and $\beta \to -\beta$) are required.
Eq. (7) tells us that both $\cal J$ and $\widetilde{\cal J}$ flip
their signs in the above replacements, and thus their ratio remains
positive.

Although Eqs. (9)---(11) are exact, they are unable to reveal the
dependence of $\widetilde{\Delta}^{}_{ij}$ on $\Delta^{}_{ij}$ in a
transparent way. It is therefore important to make reasonable
analytical approximations in this connection, so as to simplify the
relations between $\widetilde{\Delta}^{}_{ij}$ and $\Delta^{}_{ij}$.
The remarkable analytical approximations made by Freund
\cite{Freund} have been popularly applied to the studies of various
long- or medium-baseline neutrino oscillation experiments with $E
\gtrsim 0.5$ GeV
\footnote{See, also, the analytical expansions made in Refs.
\cite{Xu,Ohlsson,Cervera}. When the unitarity of the
$3\times 3$ PMNS matrix $U$ is directly or indirectly violated in
the presence of light or heavy sterile neutrinos, the similar
analytical expansions of neutrino oscillation probabilities have
been done by Li and Luo \cite{Li-Luo}.}.
Given the fact that $7.02 \times 10^{-5} ~{\rm eV}^2 \leq
\Delta^{}_{21} \leq 8.09 \times 10^{-5} ~{\rm eV}^2$ holds at the
$3\sigma$ confidence level \cite{GG} and the dependence of
terrestrial matter effects on the neutrino beam energy $E$ can be
effectively expressed as $A \simeq 2.28 \times 10^{-4} ~{\rm eV}^2
\left(E/{\rm GeV}\right)$ for a realistic ongoing or upcoming
neutrino oscillation experiment \cite{matter}
\footnote{To be more specific, the ``matter" parameter $A$ is given
as $A \simeq 1.52 \times 10^{-4} ~{\rm eV}^2 ~Y^{}_e \left(\rho/{\rm
g/cm^3}\right) \left(E/{\rm GeV}\right) \simeq 2.28 \times 10^{-4} ~
{\rm eV}^2 \left(E/{\rm GeV}\right)$, where $Y^{}_e \simeq 0.5$ is
the electron fraction and $\rho \simeq 3 ~{\rm g/cm^3}$ is the
typical matter density for a neutrino trajectory through the Earth's
crust.},
the limit $E \gtrsim 0.5$ GeV is essentially equivalent to the
requirement $|\alpha| < |\beta|$.

But we stress that the case of $|\alpha| \gtrsim |\beta|$ is also
interesting in neutrino phenomenology, especially in the aspect of
probing leptonic CP and T violation in a low-energy medium-baseline
oscillation experiment \cite{Minakata}. In fact, there will be no
way to obtain $\widetilde{\cal J}/{\cal J} \gtrsim 1$ if the
neutrino beam energy $E$ is higher than about $0.5$ GeV. To see this
point, we calculate the ratio of $\widetilde{\cal J}$ to $\cal J$ by
using Eqs. (7)---(11) and inputting the best-fit values of
$\Delta^{}_{21}$, $\Delta^{}_{31}$, $\theta^{}_{12}$ and
$\theta^{}_{13}$ listed in Table 1 \cite{GG}. Allowing $E$ to vary
from $0$ to $100$ GeV, we plot the numerical change of
$\widetilde{\cal J}/{\cal J}$ with $E$ in Fig. 1, where both the
neutrino (with $A$) and antineutrino (with $-A$) beams are
considered, together with both the normal ($\Delta^{}_{31} >0$) and
inverted ($\Delta^{}_{31} <0$) neutrino mass hierarchies. Some
observations and discussions are in order.
\begin{table}[t]
\centering \caption{The best-fit values and $3\sigma$ ranges of six
neutrino oscillation parameters from a global fit of current
experimental data \cite{GG}.} \vspace{0.4cm}
\begin{tabular}{ccccccccc}\hline
&& \multicolumn{3}{c}{Normal mass ordering (NMO)} &&
\multicolumn{3}{c}{Inverted mass ordering (IMO)} \\ \hline
&& best-fit & & $3\sigma$ range & & best-fit & & $3\sigma$ range
\\ \hline
$\theta^{}_{12}$ && $33.48^\circ$ && $31.29^\circ$ --- $35.91^\circ$
&& $33.48^\circ$ &&$31.29^\circ$ --- $35.91^\circ$ \\
$\theta^{}_{13}$ && $8.50^\circ$ && $7.85^\circ$ ---
$9.10^\circ$ &&$8.51^\circ$ && $7.87^\circ$ --- $9.11^\circ$ \\
$\theta^{}_{23}$ && $42.3^\circ$ && $38.2^\circ$ ---
$53.3^\circ$ && $49.5^\circ$ && $38.6^\circ$ --- $53.3^\circ$ \\
$\delta$ && $306^\circ$ && $0^\circ$ --- $360^\circ$ && $254^\circ$
&& $0^\circ$ --- $360^\circ$ \\
\hline \\ \vspace{-1.15cm} \\
$\displaystyle \frac{\Delta^{}_{21}}{10^{-5} ~{\rm eV}^2}$ && $7.50$
&&$7.02$ --- $8.09$ && $7.50$ && $7.02$ --- $8.09$ \\ \vspace{-0.4cm} \\
$\displaystyle \frac{\Delta^{}_{31}}{10^{-3} ~{\rm eV}^2}$ &&
$2.457$ && $2.317$ --- $2.607$ && $-2.374$ && $-2.520$ --- $-2.226$ \\
\vspace{-0.47cm} \\ \hline
\end{tabular}
\end{table}
\begin{figure*}[ht]
\vspace{0.4cm}
\centerline{\includegraphics[width=16cm]{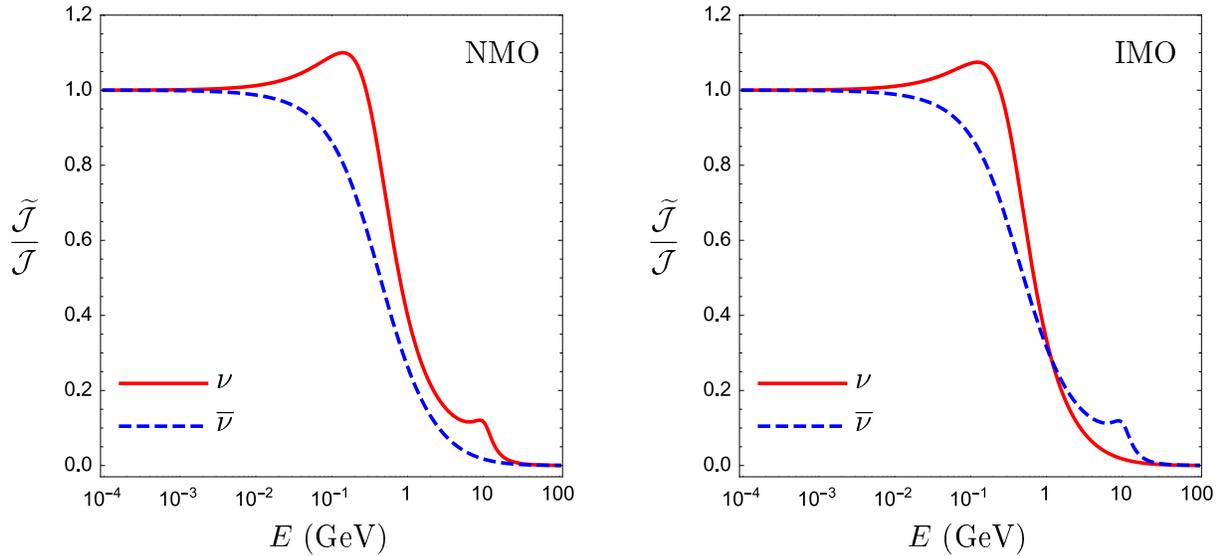}}
\caption{The ratio of the matter-corrected Jarlskog invariant
$\widetilde{\cal J}$ to its fundamental counterpart $\cal J$ as a
function of the neutrino ($\nu$ with $A$) or antineutrino
($\overline{\nu}$ with $-A$) beam energy $E$ in the case of a normal
mass ordering (NMO, left panel) or an inverted mass ordering (IMO,
right panel). Here the best-fit values of $\Delta^{}_{21}$,
$\Delta^{}_{31}$, $\theta^{}_{12}$ and $\theta^{}_{13}$ \cite{GG}
have been input.}
\end{figure*}

(1) Except the extreme case of ${\cal J} = 0$ (i.e., $\delta =0$ or
$\pi$) which makes the ratio of $\widetilde{\cal J}$ to ${\cal J}$
nonsense, the profile of $\widetilde{\cal J}/{\cal J}$ changing with
$E$ is stable and independent of the value of $\theta^{}_{23}$ and
the large uncertainties of $\delta$ itself. In all the four
possibilities shown in Fig. 1, the size of $\widetilde{\cal J}/{\cal
J}$ goes down quickly when $E$ becomes larger than about $0.5$ GeV.
As for the case of an antineutrino beam plus the normal mass
hierarchy, $\widetilde{\cal J}/{\cal J}$ decreases in a
monotonic way and does not develop any maxima or minima. In
comparison, $\widetilde{\cal J}/{\cal J}$ can have one maximum in
the case of a neutrino beam plus the inverted mass hierarchy, or one
maximum and one minimum in the case of an antineutrino beam plus the
inverted mass hierarchy, or two maxima and one minimum in the case
of a neutrino beam plus the normal mass hierarchy. But we are mainly
interested in the peaks of $\widetilde{\cal J}/{\cal J}$ in the
latter three cases, where the corresponding values of $E$ are
summarized as follows
\footnote{In the case of a neutrino (or antineutrino) beam with the
normal (or inverted) mass ordering, the minimum of $\widetilde{\cal
J}/{\cal J}$ is about $0.116$ (or $0.113$) appearing at $E \simeq
6.462$ GeV (or $6.172$ GeV). The magnitude of such an extreme is
actually similar to the suppressed peak $\widetilde{\cal
J}^\prime_*/{\cal J} \simeq 0.12$ at $E^\prime_* \simeq 8.906$ GeV
(or $8.828$ GeV).}:
\begin{eqnarray}
\nu ~ {\rm beam}~ (\Delta^{}_{31} >0): & \hspace{-0.2cm} &
\displaystyle E^{}_*
\simeq 0.140 ~{\rm GeV} \; , ~~ \frac{\widetilde{\cal J}^{}_*}{\cal
J} \simeq 1.10 \; ; ~~ E^{\prime}_* \simeq 8.906 ~{\rm GeV} \; , ~~
\frac{\widetilde{\cal J}^{\prime}_*}{\cal J} \simeq 0.12 \; ;
\nonumber \\
\nu ~ {\rm beam}~ (\Delta^{}_{31} <0): & \hspace{-0.2cm} &
\displaystyle E^{}_* \simeq 0.123 ~{\rm GeV} \; , ~~
\frac{\widetilde{\cal J}^{}_*}{\cal J} \simeq 1.07 \; ;
\nonumber \\
\overline{\nu} ~ {\rm beam}~ (\Delta^{}_{31} <0): & \hspace{-0.2cm}
& \displaystyle
E^{\prime}_* \simeq 8.828 ~{\rm GeV} \; , ~~ \frac{\widetilde{\cal
J}^{\prime}_*}{\cal J} \simeq 0.12 \; . \hspace{1cm}
\end{eqnarray}
Of course, the suppressed peaks with $\widetilde{\cal
J}^{\prime}_*/{\cal J} \simeq 0.12$ are not within the scope of our
interest in this work, because the corresponding beam energies are
far above $1$ GeV.

(2) But a suppressed peak $\widetilde{\cal J}^{\prime}_*/{\cal J}
\simeq 0.12$ and its resonance energy $E^\prime_*$ can be well
understood by following the analytical approximations made in Ref.
\cite{Freund} for $E \gtrsim 1$ GeV. Namely,
\begin{eqnarray}
\frac{\widetilde{\cal J}}{\cal J} \simeq
\frac{\Delta^{}_{21}}{\displaystyle A \cos^2 \theta^{}_{13}
\sqrt{\displaystyle \beta^2 - 2 \beta \cos 2\theta^{}_{13} + 1}}
\;\; ,
\end{eqnarray}
in which $\beta = \pm A/\Delta^{}_{31}$ correspond
to the neutrino and antineutrino beams, respectively.
We find that this ratio peaks at
\begin{eqnarray}
\beta^{\prime}_* = \frac{\pm A^\prime_*}{\Delta^{}_{31}}
\simeq \frac{3 \cos 2\theta^{}_{13} +
\sqrt{\displaystyle 1 - 9 \sin^2 2\theta^{}_{13}}}{4} \;
\end{eqnarray}
with $A^\prime_* = 2\sqrt{2} \ G^{}_{\rm F} N^{}_e E^\prime_*$,
where the smallness of $\theta^{}_{13}$ has been taken into account.
Now that $\beta^\prime_*$ itself is positive, the plus (or minus)
sign in front of $A^\prime_*$ in Eq. (14) must correspond to the
neutrino (or antineutrino) beam with the normal (or inverted) mass
ordering. Given the best-fit value of $\theta^{}_{13}$ in Table 1,
it is straightforward to obtain $E^\prime_* \simeq 9.02$ GeV in the
$\Delta^{}_{13} >0$ case or $E^\prime_* \simeq 8.71$ GeV in the
$\Delta^{}_{13} <0$ case. Such approximate results are in agreement
with the exact numerical results shown in Eq. (12) to a reasonably
good degree of accuracy.
\begin{figure*}[t]
\vspace{0.4cm}
\centerline{\includegraphics[width=16cm]{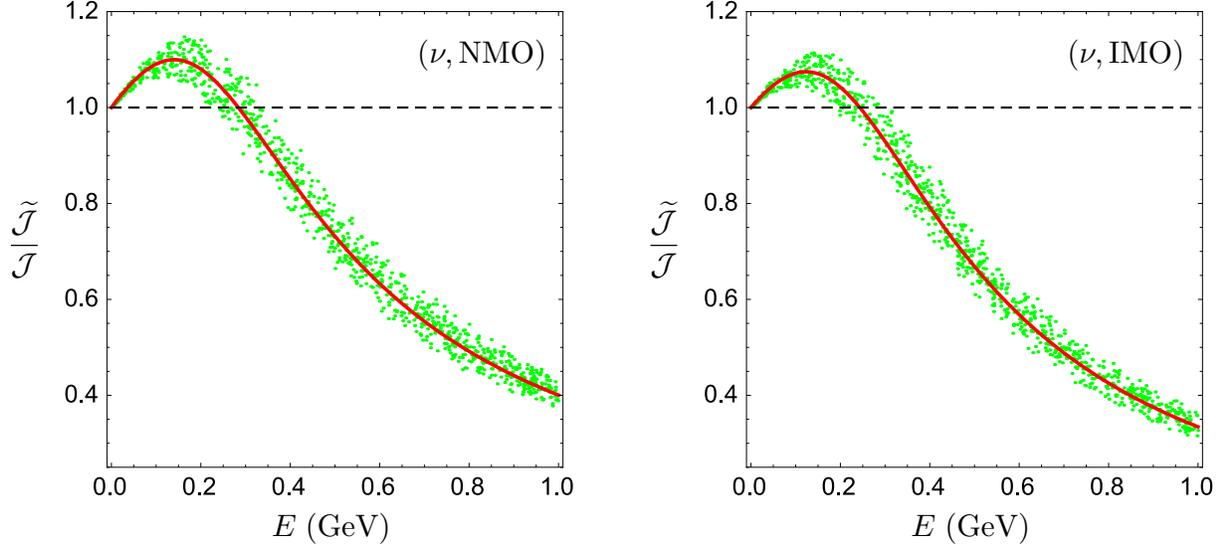}}
\caption{The ratio $\widetilde{\cal J}/{\cal J}$ as a function of
the {\it neutrino} beam energy $E$ in the normal or inverted mass
ordering case. Here the red curve and the green band correspond to
the inputs of the best-fit values and $3\sigma$ ranges of
$\Delta^{}_{21}$, $\Delta^{}_{31}$, $\theta^{}_{12}$ and
$\theta^{}_{13}$ \cite{GG}, respectively.}
\end{figure*}

From now on let us concentrate on the first (matter-enhanced) peak
$\widetilde{\cal J}^{}_*/{\cal J} >1$ around $E^{}_* \in (0.1, 0.2)$
GeV in Fig. 1 and understand why it appears in an approximate but
more transparent way. Fig. 2 is a clearer illustration of this peak,
where the $3\sigma$ ranges of $\Delta^{}_{21}$, $\Delta^{}_{31}$,
$\theta^{}_{12}$ and $\theta^{}_{13}$ are also taken into account.
One can see that the numerical uncertainties associated with the
four input parameters do not change the lineshape of
$\widetilde{\cal J}/{\cal J}$, implying that our analytical
approximations to be made below will keep valid when the relevant
neutrino oscillation parameters are measured to a much higher degree
of accuracy in the near future. In the low-energy region under
consideration the magnitude of $\beta$ is comparable with or smaller
than that of $\alpha$, and thus both of them can serve for the small
expansion parameters in our analytical approximations for
$\widetilde{\Delta}^{}_{21}$, $\widetilde{\Delta}^{}_{31}$ and
$\widetilde{\Delta}^{}_{32}$. We first consider the {\it neutrino}
beam. A tedious but straightforward calculation leads us to the
results
\begin{eqnarray}
&& \sqrt{\displaystyle x^2 - 3y} \ \simeq
\Delta^{}_{31} \left[ 1 - \frac{1}{2} \alpha -
\frac{1}{2} \left(1 - 3|U^{}_{e 3}|^2\right) \beta + \frac{3}{8}
\alpha^2 -\frac{3}{4} \left(|U^{}_{e1}|^2 - |U^{}_{e2}|^2\right)
\alpha\beta \right. \hspace{0.8cm}
\nonumber \\
&& \hspace{3.2cm} \left.
+ \frac{3}{8} \left(1 + 2|U^{}_{e 3}|^2\right) \beta^2 \right] \; ,
\nonumber \\
&& \displaystyle z \simeq 1 - \frac{3}{8} \alpha^2 + \frac{3}{4}
\left(|U^{}_{e 1}|^2 - |U^{}_{e 2}|^2\right) \alpha\beta -
\frac{3}{8} \left(1 - 2|U^{}_{e 3}|^2\right) \beta^2 \; ,
\nonumber \\
&& \sqrt{\displaystyle 3 \left(1 - z^2\right)}
\simeq \frac{3}{2} \epsilon \left(1 + \frac12 \alpha +
\frac12 \beta \right) \;
\end{eqnarray}
for the $\Delta^{}_{31} >0$ case; and
\begin{eqnarray}
&& \sqrt{\displaystyle x^2 - 3y} \simeq
-\Delta^{}_{31} \left[ 1 - \frac{1}{2} \alpha -
\frac{1}{2} \left(1 - 3|U^{}_{e 3}|^2\right) \beta + \frac{3}{8}
\alpha^2 -\frac{3}{4} \left(|U^{}_{e1}|^2 - |U^{}_{e2}|^2\right)
\alpha\beta \right. \hspace{0.7cm}
\nonumber \\
&& \hspace{3.3cm} \left.
+ \frac{3}{8} \left(1 + 2|U^{}_{e 3}|^2\right) \beta^2 \right] \;
,\nonumber \\
&& z \simeq \frac{1}{2} + \frac{3}{4} \epsilon \left(1 + \frac12
\alpha + \frac12 \beta \right)- \frac{3}{16}
\alpha^2 + \frac{3}{8} \left(|U^{}_{e 1}|^2 - |U^{}_{e 2}|
^2\right)\alpha\beta - \frac{3}{16} \left(1 - 2|U^{}_{e 3}|
^2\right) \beta^2\; ,
\nonumber \\
&& \sqrt{\displaystyle 3 \left(1 - z^2\right)} \simeq
\frac{3}{2} - \frac{3}{4} \epsilon \left(1 + \frac12 \alpha +
\frac12 \beta \right) - \frac{9}{16} \alpha^2
+ \frac{9}{8} \left(|U^{}_{e 1}|^2 - |U^{}_{e 2}|^2\right)
\alpha\beta \nonumber \\
&& \hspace{2.75cm} - \frac{9}{16} \left(1 - 2|U^{}_{e 3}|^2\right)
\beta^2 \;
\end{eqnarray}
for the $\Delta^{}_{31} <0$ case, where
\begin{eqnarray}
\epsilon \equiv \sqrt{\displaystyle \alpha^2 - 2 \left(|U^{}_{e
1}|^2 - |U^{}_{e 2}|^2 \right) \alpha\beta + \left(1 - |U^{}_{e
3}|^2 \right)^2 \beta^2} \;
\end{eqnarray}
is a small parameter, and the smallness of $|U^{}_{e 3}|$ is already
implied. Then we obtain the effective neutrino mass-squared
differences from Eq. (9) or Eq. (10):
\begin{eqnarray}
\widetilde{\Delta}^{}_{21} \hspace{-0.15cm} & \simeq &
\hspace{-0.15cm} \Delta^{}_{31} \left(1 +
\frac{3}{2} |U^{}_{e 3}|^2 \beta \right) \epsilon \; ,
\nonumber \\
\widetilde{\Delta}^{}_{31} \hspace{-0.15cm} & \simeq &
\hspace{-0.15cm} \Delta^{}_{31} \left[1 - \frac{1}{2} \alpha -
\frac{1}{2} \left(1 - 3|U^{}_{e 3}|^2 \right) \beta + \frac{1}{2}
\epsilon +\frac{3}{4} |U^{}_{e 3}|^2 \beta \epsilon + \frac
32 |U^{}_{e 3}|^2 \beta^2\right]
\; ,
\nonumber \\
\widetilde{\Delta}^{}_{32} \hspace{-0.15cm} & \simeq &
\hspace{-0.15cm} \Delta^{}_{31} \left[1 - \frac{1}{2} \alpha -
\frac{1}{2} \left(1 - 3|U^{}_{e 3}|^2 \right) \beta - \frac{1}{2}
\epsilon  - \frac{3}{4} |U^{}_{e 3}|^2 \beta \epsilon + \frac
32 |U^{}_{e 3}|^2 \beta^2 \right]
\; ,
\end{eqnarray}
for the $\Delta^{}_{31} >0$ case; or
\begin{eqnarray}
\widetilde{\Delta}^{}_{21} \hspace{-0.15cm} & \simeq &
\hspace{-0.15cm} -\Delta^{}_{31} \left(1 + \frac{3}{2} |U^{}
_{e 3}|^2 \beta  \right) \epsilon \; ,
\nonumber \\
\widetilde{\Delta}^{}_{31} \hspace{-0.15cm} & \simeq &
\hspace{-0.15cm} \Delta^{}_{31} \left[1 - \frac{1}{2} \alpha -
\frac{1}{2} \left(1 - 3|U^{}_{e 3}|^2 \right) \beta - \frac{1}{2}
\epsilon - \frac{3}{4} |U^{}_{e 3}|^2 \beta \epsilon + \frac
32 |U^{}_{e 3}|^2 \beta^2  \right]
\; ,
\nonumber \\
\widetilde{\Delta}^{}_{32} \hspace{-0.15cm} & \simeq &
\hspace{-0.15cm} \Delta^{}_{31} \left[1 - \frac{1}{2} \alpha -
\frac{1}{2} \left(1 - 3|U^{}_{e 3}|^2 \right) \beta + \frac{1}{2}
\epsilon + \frac{3}{4} |U^{}_{e 3}|^2 \beta \epsilon + \frac
32 |U^{}_{e 3}|^2 \beta^2  \right]
\; ,
\end{eqnarray}
for the $\Delta^{}_{31} <0$ case. Given the standard parametrization
of the PMNS mixing matrix $U$ in Eq. (8), the small parameter
$\epsilon$ in Eq. (17) can be reexpressed as
\begin{eqnarray}
\epsilon = \sqrt{\displaystyle \alpha^2 - 2 \alpha\beta
\cos 2\theta^{}_{12} \cos^2 \theta^{}_{13} +
\beta^2 \cos^4 \theta^{}_{13}} \;\; ,
\end{eqnarray}
so its magnitude is apparently of ${\cal O}(\alpha)$ or ${\cal
O}(\beta)$. With the help of Eqs. (7), (18) and (19), we arrive at the
approximate analytical results for the ratio of $\widetilde{\cal J}$
to $\cal J$ as follows:
\begin{eqnarray}
\frac{\widetilde{\cal J}}{\cal J} \simeq +\frac{\alpha}{\epsilon}
\left(1 + \beta\right) \simeq \frac{\alpha}{\displaystyle
\sqrt{\displaystyle \alpha^2 - 2 \alpha\beta \cos 2\theta^{}_{12} +
\beta^2}} \left(1 + \beta\right) \; ,
\end{eqnarray}
for the $\Delta^{}_{31} >0$ case; or
\begin{eqnarray}
\frac{\widetilde{\cal J}}{\cal J} \simeq -\frac{\alpha}{\epsilon}
\left(1 + \beta\right) \simeq \frac{-\alpha}{\displaystyle
\sqrt{\displaystyle \alpha^2 - 2 \alpha\beta \cos 2\theta^{}_{12} +
\beta^2}} \left(1 +\beta\right) \; ,
\end{eqnarray}
for the $\Delta^{}_{31} <0$ case, in which the terms proportional to
$|U^{}_{e 3}|^2 = \sin^2 \theta^{}_{13}$ in $\epsilon$ have been
omitted thanks to the smallness of $\theta^{}_{13}$. Since
$\epsilon$ has a minimum value $\epsilon^{}_* \simeq |\alpha| \sin
2\theta^{}_{12}$ at $\beta^{}_* \simeq \alpha \cos 2\theta^{}_{12}$,
we expect that the ratio $\widetilde{\cal J}/{\cal J}$ takes its
maximum value $1/\sin 2\theta^{}_{12}$ in the leading-order
approximation, no matter whether the neutrino mass ordering is
normal or inverted. As for an {\it antineutrino} beam, the matter
parameter is actually $-A$, and thus the replacement $\beta \to
-\beta$ must be made for the analytical results obtained above. In
other words, $\epsilon$ does not develop a minimum value in the {\it
antineutrino} case --- that is why $\widetilde{\cal J}/{\cal J}$
does not undergo any resonances in this case, a conclusion
independent of the neutrino mass ordering. So we only concentrate on
the {\it neutrino} beam in the subsequent discussions.

Let us go beyond the leading-order approximation to calculate the
extreme value of $\widetilde{\cal J}/{\cal J}$, which is a function
of $\beta$ (or equivalently, the matter parameter $A$ or the
neutrino beam energy $E$). To do so, we take the first derivative of
$\widetilde{\cal J}/{\cal J}$ with respect to the variable $\beta$
in Eq. (21) or (22) and set it to equal zero, and find that such a
treatment leads to the same equation in these two cases:
\begin{eqnarray}
\left[ \left(1 - |U^{}_{e 3}|^2\right)^2 + \alpha \left(|U^{}
_{e 1}|^2 - |U^{}_{e 2}|^2 \right)\right] \beta - \left(|U^{}
_{e 1}|^2 - |U^{}_{e 2}|^2\right)\alpha - \alpha^2 \simeq 0 \; .
\end{eqnarray}
The solution to Eq. (23) turns out to be
\begin{eqnarray}
\beta^{}_* = \frac{A^{}_*}{\Delta^{}_{31}} \simeq \alpha \left[
\cos 2\theta^{}_{12} \left(1 + \sin^2\theta^{}_{13}\right)
+  \alpha \sin^2 2\theta^{}_{12} \right] \;
\end{eqnarray}
with $A^{}_* = 2\sqrt{2} \ G^{}_{\rm F} N^{}_e E^{}_*$, from which
one can easily obtain the resonance energy $E^{}_*$ that has been
given in Eq. (3). Substituting Eq. (24) into Eq. (21) or (22), we
immediately arrive at the maximum value of $\widetilde{\cal J}/{\cal
J}$ on the resonance:
\begin{eqnarray}
\frac{\widetilde{\cal J}^{}_*}{\cal J} \hspace{-0.15cm} & \simeq &
\hspace{-0.15cm} \frac{1}{\sin 2\theta^{}_{12}} \left[ 1 +
\alpha \displaystyle \cos 2\theta^{}_{12}
\left( 1 + \sin^2\theta^{}_{13}\right)
 + \frac{1}{2} \alpha^2 \sin^2 2\theta^{}_{12}
\right] \; ,
\end{eqnarray}
an interesting and instructive result whose leading and
next-to-leading-order parts have been shown in Eq. (4). Taking the
best-fit values of $\theta^{}_{12}$, $\theta^{}_{13}$,
$\Delta^{}_{21}$ and $\Delta^{}_{31}$ for example, we obtain $E^{}_*
\simeq 0.140$ GeV (or $0.123$ GeV) and $\widetilde{\cal
J}^{}_*/{\cal J} \simeq 1.10$ (or $1.07$) for the normal (or
inverted) neutrino mass ordering from the analytical formulas in
Eqs. (24) and (25), in good agreement with the more exact numerical
results that have been listed in Eq. (12).
\begin{figure*}[t]
\vspace{0.4cm}
\centerline{\includegraphics[width=15.6cm]{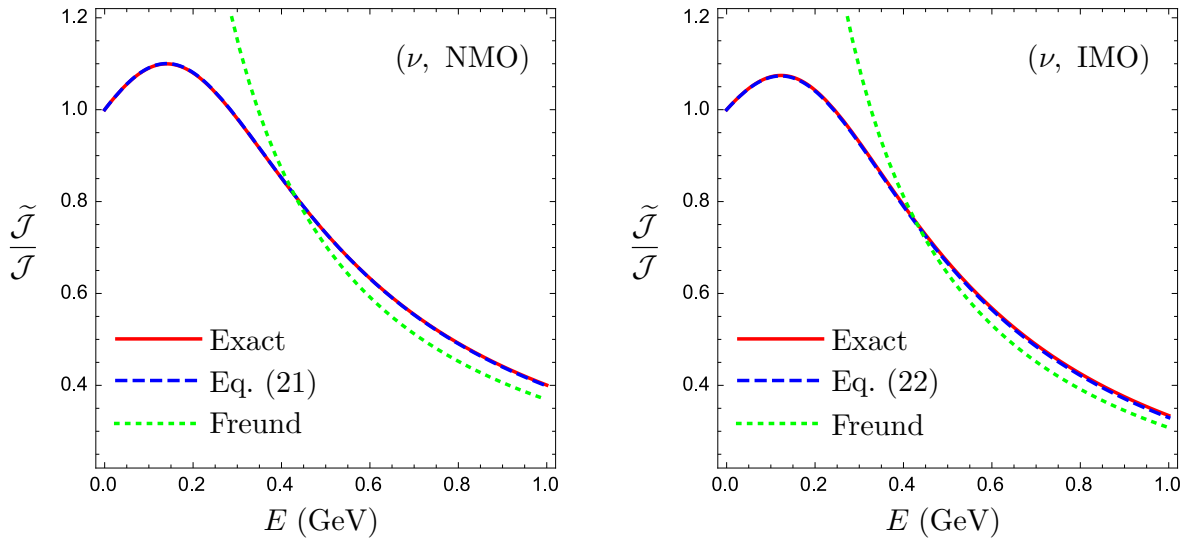}}
\caption{A numerical comparison between the results of
$\widetilde{\cal J}/{\cal J}$ obtained from the exact formula in Eq.
(7) (red line) and our analytical approximations in Eqs. (21) and
(22) (blue dashed curve) or Freund's approximation in Eq. (13)
(green dotted curve), where $E$ varies from zero to $1$ GeV, and the
best-fit values of $\Delta^{}_{21}$, $\Delta^{}_{31}$,
$\theta^{}_{12}$ and $\theta^{}_{13}$ \cite{GG} have been input.}
\end{figure*}

In Fig. 3 we compare the result of $\widetilde{\cal J}/{\cal J}$
obtained from our analytical approximation made in Eq. (21) or (22)
with its exact numerical result by allowing the neutrino beam energy
$E$ to vary from zero to $1$ GeV. We see that the two sets of
results agree with each other in a perfect way. In contrast, the
numerical result of $\widetilde{\cal J}/{\cal J}$ obtained from
Freund's analytical approximation in Eq. (13) is not so good in the
$0.4 ~{\rm GeV} \lesssim E \lesssim 1 ~{\rm GeV}$ range, and it
becomes out of control for $E \lesssim 0.4$ GeV. Hence our
analytical approximations stand out as a much better tool of
understanding the salient features of the matter-corrected Jarlskog
parameter in the $E \lesssim 1$ GeV region. In fact, the typical
neutrino beam energy of the realistic T2K long-baseline oscillation
experiment \cite{T2K} is about $0.6$ GeV, just within this region.
So one may use the analytical formulas given in the present work to
do a reliable phenomenological analysis of CP violation and the
associated matter effects in the T2K experiment.

Given the resonance energy $E^{}_*$ in Eq. (3) and the maximum value
$\widetilde{\cal J}^{}_*/{\cal J}$ in Eq. (4), the profiles of
$\widetilde{\cal J}/{\cal J}$ in the left and right panels of Fig. 2
can easily be understood. Simply because the next-to-leading-order
terms of $E^{}_*$ and $\widetilde{\cal J}^{}_*/{\cal J}$ are both
proportional to the expansion parameter $\alpha =
\Delta^{}_{21}/\Delta^{}_{31} \simeq \pm 1/30$, they exhibit a small
but appreciable difference in Fig. 2 with respect to the normal and
inverted neutrino mass hierarchies. This observation indicates that
even a low-energy neutrino oscillation experiment could have the
potential to probe not only the CP- and T-violating effects but also
the neutrino mass ordering.

At this point it is worth stressing that the matter-induced
amplification or enhancement of $\widetilde{\cal J}$ under
discussion is actually associated with the sensitivity of
$\theta^{}_{12}$ to the matter-induced correction. It is well known
that $\theta^{}_{13}$, $\theta^{}_{23}$ and $\delta$ are almost
insensitive to terrestrial matter effects (i.e.,
$\widetilde{\theta}^{}_{13} \simeq \theta^{}_{13}$,
$\widetilde{\theta}^{}_{23} \simeq \theta^{}_{23}$ and
$\widetilde{\delta} \simeq \delta$) in the $E \lesssim 1$ GeV region
\cite{Ohlsson,Minakata,Xing04}, and hence the first equality in Eq.
(7) leads us to the relation
\begin{eqnarray}
\frac{\sin 2\widetilde{\theta}^{}_{12}}{\sin 2\theta^{}_{12}}
\simeq \frac{\widetilde{\cal J}}{\cal J} \simeq
\frac{|\alpha|}{\epsilon} \left(1 + \beta\right) \; ,
\end{eqnarray}
where Eq. (21) or (22) has been taken into account. So the behavior
of the ratio of $\sin 2\widetilde{\theta}^{}_{12}$ to $\sin
2\theta^{}_{12}$ changing with $E$ is expected to be the same as
that of $\widetilde{\cal J}/{\cal J}$ shown in Fig. 2.

Last but not least, let us figure out the upper limit of $E$ which
allows $\widetilde{\cal J}/{\cal J} \gtrsim 1$ to hold. For this
purpose, we take $\widetilde{\cal J}/{\cal J} \simeq 1$ in Eq. (21)
or (22) and then solve this equation. Besides the trivial solution
$E =0$, there is a nontrivial solution
\begin{eqnarray}
E^{}_0 \simeq \frac{\Delta^{}_{21}}{ \sqrt{2} \ G^{}_{\rm F}
N^{}_e} \left[ \cos 2\theta^{}_{12} \left(1 +\sin^2 \theta_{13}^{}
\right) + \alpha \right] \; ,
\end{eqnarray}
which is valid for both normal and inverted neutrino mass
hierarchies. Namely, $\widetilde{\cal J}/{\cal J} \gtrsim 1$ holds
for $E \in \left[0, E^{}_0\right]$ --- the region of $E$ which might
be especially interesting for the study of leptonic CP violation in
a low-energy medium-baseline neutrino oscillation experiment. If
only the leading term in Eq. (27) is taken into account (i.e.,
omitting the $\alpha$ term and taking $\sin^2 \theta^{}_{13}
\simeq 0$), we are then left with $E^{}_0 \simeq \Delta^{}_{21} \cos
2\theta^{}_{12}/ \left(\sqrt{2} \ G^{}_{\rm F} N^{}_e\right) \lesssim
0.3$ GeV by considering $A \simeq 2.28 \times 10^{-4} ~{\rm eV}^2
\left(E/{\rm GeV} \right)$ and inputting the best-fit values of
$\Delta^{}_{21}$ and $\theta^{}_{12}$ \cite{GG}. Given the best-fit
values and $3\sigma$ ranges of $\Delta^{}_{21}$, $\Delta^{}_{31}$,
$\theta^{}_{12}$ and $\theta^{}_{13}$ listed in Table 1, the more
accurate results of $E^{}_0$ can be obtained from solving
$\widetilde{\cal J}/{\cal J} =1$ in Eq. (7) in a numerical way:
$E^{}_0 \simeq 0.284$ GeV (best-fit) and $0.214 ~{\rm GeV} \lesssim
E^{}_0 \lesssim 0.359 ~{\rm GeV}$ ($3\sigma$ range) for the normal
neutrino mass ordering, or $E^{}_0 \simeq 0.244$ GeV (best-fit) and
$0.179 ~{\rm GeV} \lesssim E^{}_0 \lesssim 0.312 ~{\rm GeV}$
($3\sigma$ range) for the inverted neutrino mass ordering. These
results are consistent with those shown in Fig. 2.

Since $E^{}_0 \simeq 2 E^{}_*$ holds as a good approximation, one
could consider to set the neutrino beam energy $E$ in the $E^{}_*
\lesssim E \lesssim 2 E^{}_*$ range when designing a realistic
medium-baseline oscillation experiment to probe the $\widetilde{\cal
J}/{\cal J} \gtrsim 1$ region of CP violation. In fact, the typical
beam energies of the proposed MOMENT \cite{MOMENT} and ESS$\nu$SM
\cite{ESS} experiments just lie in such an interesting region.

\section{The matter-deformed unitarity triangles}

The three Dirac UTs defined in vacuum in Eq. (5) have their
counterparts in matter, namely,
\begin{eqnarray}
\widetilde{\triangle}^{}_e : && \widetilde{U}^{}_{\mu 1}
\widetilde{U}^*_{\tau 1} + \widetilde{U}^{}_{\mu 2}
\widetilde{U}^*_{\tau 2} + \widetilde{U}^{}_{\mu 3}
\widetilde{U}^*_{\tau 3} = 0 \; ,
\nonumber \\
\widetilde{\triangle}^{}_\mu : && \widetilde{U}^{}_{\tau 1}
\widetilde{U}^*_{e 1} + \widetilde{U}^{}_{\tau 2}
\widetilde{U}^*_{e 2} + \widetilde{U}^{}_{\tau 3}
\widetilde{U}^*_{e 3} = 0 \; ,
\nonumber \\
\widetilde{\triangle}^{}_\tau : && \widetilde{U}^{}_{e 1}
\widetilde{U}^*_{\mu 1} + \widetilde{U}^{}_{e 2}
\widetilde{U}^*_{\mu 2} + \widetilde{U}^{}_{e 3}
\widetilde{U}^*_{\mu 3} = 0 \; . \hspace{0.7cm}
\end{eqnarray}
Thanks to the unitarity of $U$ and $\widetilde{U}$, the areas of
$\triangle^{}_\alpha$ and $\widetilde{\triangle}^{}_\alpha$ (for
$\alpha = e, \mu, \tau$) are equal to $|{\cal J}|/2$ and
$|\widetilde{\cal J}|/2$, respectively. Hence a change of the ratio
$\widetilde{\cal J}/{\cal J}$ with the neutrino beam energy $E$
implies that the three UTs must be deformed by terrestrial matter
effects. The exact analytical expressions of the three sides of
$\widetilde{\triangle}^{}_\alpha$ for a constant matter profile have
been derived in Ref. \cite{Zhang}. Here we find a more convenient
way to reexpress the previous results
\footnote{In the low-energy region under consideration we find that
the $U^{}_{\alpha 3} U^*_{\beta 3}$ side of $\triangle^{}_{\gamma}$,
where the subscripts $\alpha$, $\beta$ and $\gamma$ run over $e$,
$\mu$ and $\tau$ cyclically, is least sensitive to terrestrial
matter effects. Hence it is appropriate to express the other two
sides in matter as $\widetilde{U}^{}_{\alpha i}
\widetilde{U}^*_{\beta i} = c^{}_1 U^{}_{\alpha i} U^*_{\beta i} +
c^{}_2 U^{}_{\alpha 3} U^*_{\beta 3}$ (for $i = 1$ or $2$), in which
the coefficients $c^{}_1$ and $c^{}_2$ deviate respectively from $1$
and $0$ due to the matter-induced corrections.},
and take into account both normal and inverted neutrino mass
hierarchies
\footnote{In this connection only the possibility of a normal
neutrino mass hierarchy is discussed analytically and numerically in
the literature. The present work improves the previous ones by
taking account of both normal and inverted mass hierarchies and shows
the phenomenological differences between these two cases.}.
To be specific, we obtain the formulas for a {\it neutrino} beam as
follows:
\begin{eqnarray}
\widetilde{\triangle}_e : ~ \left\{
\begin{array}{lcl} \widetilde{U}^{}_{\mu
1}\widetilde{U}_{\tau 1}^{\ast} \hspace{-0.15cm} & = &
\hspace{-0.15cm} \displaystyle \frac{\left(\Delta^{\prime}_{31} +
A\right) \Delta^{}_{21}}{\widetilde{\Delta}^{}_{21}
\widetilde{\Delta}^{}_{31}} U^{}_{\mu 1} U_{\tau 1}^{\ast} -
\frac{\left(\Delta^{\prime}_{11} + A\right) \Delta^{}_{32}}
{\widetilde{\Delta}^{}_{21} \widetilde{\Delta}^{}_{31}} U^{}_{\mu 3}
U_{\tau 3}^{\ast} \; ,
\\ \vspace{-0.4cm} \\
\widetilde{U}^{}_{\mu
2}\widetilde{U}_{\tau 2}^{\ast} \hspace{-0.15cm} & = &
\hspace{-0.15cm}  \displaystyle
\frac{\left(\Delta^{\prime}_{32}+A\right)
\Delta^{}_{21}}{\widetilde{\Delta}^{}_{21}
\widetilde{\Delta}^{}_{32}} U^{}_{\mu 2} U_{\tau 2}^{\ast} +
\frac{\left(\Delta^{\prime}_{22}+A\right)
\Delta^{}_{31}}{\widetilde{\Delta}^{}_{21}
\widetilde{\Delta}^{}_{32}} U^{}_{\mu 3} U_{\tau 3}^{\ast} \; ,
\\ \vspace{-0.4cm} \\
\widetilde{U}^{}_{\mu 3}\widetilde{U}_{\tau 3}^{\ast}
\hspace{-0.15cm} & = & \hspace{-0.15cm}  \displaystyle
\frac{\left(\Delta^{\prime}_{13}+A\right)\Delta^{}_{23}}
{\widetilde{\Delta}^{}_{31} \widetilde{\Delta}^{}_{32}} U^{}_{\mu
3}U_{\tau 3}^{\ast} + \frac{\left(\Delta^{\prime}_{33}+A\right)
\Delta^{}_{21}}{\widetilde{\Delta}^{}_{31}
\widetilde{\Delta}^{}_{32}} U^{}_{\mu 1} U_{\tau 1}^{\ast} \; ;
\end{array} \right.
\end{eqnarray}
and
\begin{eqnarray}
\widetilde{\triangle}_\mu : ~ \left\{
\begin{array}{lcl}
\widetilde{U}^{}_{\tau 1}\widetilde{U}_{e 1}^{\ast} \hspace{-0.15cm}
& = & \hspace{-0.15cm} \displaystyle
\frac{\Delta^{\prime}_{31}\Delta^{}_{21}}{\widetilde{\Delta}^{}_{21}
\widetilde{\Delta}^{}_{31}} U^{}_{\tau 1} U_{e 1}^{\ast} -
\frac{\Delta^{\prime}_{11}
\Delta^{}_{32}}{\widetilde{\Delta}^{}_{21}\widetilde{\Delta}^{}_{31}}
U^{}_{\tau 3} U_{e 3}^{\ast} \; ,
\\ \vspace{-0.4cm} \\
\widetilde{U}^{}_{\tau 2}\widetilde{U}_{e 2}^{\ast} \hspace{-0.15cm}
& = & \hspace{-0.15cm} \displaystyle
\frac{\Delta^{\prime}_{32}\Delta^{}_{21}}{\widetilde{\Delta}^{}_{21}
\widetilde{\Delta}^{}_{32}} U^{}_{\tau 2} U_{e 2}^{\ast}
+\frac{\Delta^{\prime}_{22}
\Delta^{}_{31}}{\widetilde{\Delta}^{}_{21}\widetilde{\Delta}^{}_{32}}
U^{}_{\tau 3} U_{e 3}^{\ast} \; ,
\\ \vspace{-0.4cm} \\
\widetilde{U}^{}_{\tau 3}\widetilde{U}_{e 3}^{\ast} \hspace{-0.15cm}
& = & \hspace{-0.15cm} \displaystyle
\frac{\Delta^{\prime}_{13}\Delta^{}_{23}}{\widetilde{\Delta}^{}_{31}
\widetilde{\Delta}^{}_{32}} U^{}_{\tau 3} U_{e 3}^{\ast}
+\frac{\Delta^{\prime}_{33}
\Delta^{}_{21}}{\widetilde{\Delta}^{}_{31}\widetilde{\Delta}^{}_{32}}
U^{}_{\tau 1} U_{e 1}^{\ast} \; ;
\end{array} \right.
\end{eqnarray}
and
\begin{eqnarray}
\widetilde{\triangle}_\tau : ~ \left\{
\begin{array}{lcl}
\widetilde{U}^{}_{e1}\widetilde{U}_{\mu 1}^{\ast} \hspace{-0.15cm} &
= & \hspace{-0.15cm} \displaystyle \frac{\Delta^{\prime}_{31}
\Delta^{}_{21}}{\widetilde{\Delta}^{}_{21}
\widetilde{\Delta}^{}_{31}} U^{}_{e1} U_{\mu 1}^{\ast} -
\frac{\Delta^{\prime}_{11}
\Delta^{}_{32}}{\widetilde{\Delta}^{}_{21}\widetilde{\Delta}^{}_{31}}
U^{}_{e3} U_{\mu 3}^{\ast} \; ,
\\ \vspace{-0.4cm} \\
\widetilde{U}^{}_{e2}\widetilde{U}_{\mu 2}^{\ast} \hspace{-0.15cm} &
= & \hspace{-0.15cm} \displaystyle
\frac{\Delta^{\prime}_{32}\Delta^{}_{21}}{\widetilde{\Delta}^{}_{21}
\widetilde{\Delta}^{}_{32}} U^{}_{e2} U_{\mu 2}^{\ast} +
\frac{\Delta^{\prime}_{22}
\Delta^{}_{31}}{\widetilde{\Delta}^{}_{21}\widetilde{\Delta}^{}_{32}}
U^{}_{e3} U_{\mu 3}^{\ast} \; ,
\\ \vspace{-0.4cm} \\
\widetilde{U}^{}_{e3}\widetilde{U}_{\mu 3}^{\ast} \hspace{-0.15cm} &
= & \hspace{-0.15cm} \displaystyle
\frac{\Delta^{\prime}_{13}\Delta^{}_{23}}{\widetilde{\Delta}^{}_{31}
\widetilde{\Delta}^{}_{32}} U^{}_{e3} U_{\mu 3}^{\ast} +
\frac{\Delta^{\prime}_{33}
\Delta^{}_{21}}{\widetilde{\Delta}^{}_{31}\widetilde{\Delta}^{}_{32}}
U^{}_{e1} U_{\mu 1}^{\ast} \; ,
\end{array} \right.
\end{eqnarray}
where $\Delta^{}_{ij} \equiv m^2_i - m^2_j$,
$\widetilde{\Delta}^{}_{ij} \equiv \widetilde{m}^2_i -
\widetilde{m}^2_j$ and $\Delta^{\prime}_{ij} \equiv m^2_i -
\widetilde{m}^2_j = \Delta^{}_{ij} + \Delta^{\prime}_{jj}$ (for $i,j
= 1,2,3$) with $\Delta^{\prime}_{jj}$ being expressed as
\begin{eqnarray}
\Delta^{\prime}_{11} \hspace{-0.15cm} & = & \hspace{-0.15cm}
\displaystyle -\frac{1}{3} x +
\frac{1}{3} \sqrt{x^2 - 3y} \left [ z + \sqrt{3 \left (1 - z^2
\right)} \right ] \; ,
\nonumber \\
\Delta^{\prime}_{22} \hspace{-0.15cm} & = & \hspace{-0.15cm}
\displaystyle -\frac{1}{3} x +
\frac{1}{3} \sqrt{x^2 - 3y} \left [ z - \sqrt{3 \left (1 - z^2
\right)} \right ] + \Delta^{}_{21} \; ,
\nonumber \\
\Delta^{\prime}_{33} \hspace{-0.15cm} & = & \hspace{-0.15cm}
\displaystyle -\frac{1}{3} x - \frac{2}{3} z \sqrt{x^2 - 3y} \ +
\Delta^{}_{31} \;
\end{eqnarray}
in the $\Delta^{}_{31} >0$ case; or
\begin{eqnarray}
\Delta_{11}^{\prime} \hspace{-0.15cm} & = & \hspace{-0.15cm}
\displaystyle -\frac{1}{3} x + \frac{1}{3} \sqrt{x^2 -3y} \Big[z -
\sqrt{3\left(1-z^2\right)}\Big] \; ,
\nonumber \\
\Delta_{22}^{\prime} \hspace{-0.15cm} & = & \hspace{-0.15cm}
\displaystyle -\frac{1}{3} x - \frac{2}{3} z \sqrt{x^2 - 3y} \ +
\Delta_{21}^{} \; ,
\nonumber \\
\Delta^{\prime}_{33} \hspace{-0.15cm} & = & \hspace{-0.15cm}
\displaystyle -\frac{1}{3} x + \frac{1}{3} \sqrt{x^2 - 3y} \Big[z +
\sqrt{3\left(1-z^2\right)}\Big] + \Delta_{31}^{} \; \hspace{0.28cm}
\end{eqnarray}
in the $\Delta^{}_{31} <0$ case. Eqs. (29)---(33) are exact and
elegant in showing the corrections of terrestrial matter to the
sides of three UTs, but they are unable to give one a ball-park
feeling of the order of magnitude of such corrections due to the
complication of $\widetilde{\Delta}^{}_{ij}$ and
$\Delta^{\prime}_{ij}$. Hence it is necessary to make some
analytical approximations in order to show the deviation of
$\widetilde{\triangle}^{}_\alpha$ from $\triangle^{}_\alpha$ (for
$\alpha = e, \mu, \tau$) in a direct and transparent way.

Note that such an exercise is not only conceptually interesting and
intuitive but also helpful for expanding the matter-corrected
probabilities of $\nu^{}_\mu \to \nu^{}_e$ and
$\overline{\nu}^{}_\mu \to \overline{\nu}^{}_e$ oscillations in
terms of the small parameters $\alpha$ and $\beta$ in the $E
\lesssim 1$ GeV region, because the CP-conserving parts of
$\widetilde{P}(\nu^{}_\mu \to \nu^{}_e)$ and
$\widetilde{P}(\overline{\nu}^{}_\mu \to \overline{\nu}^{}_e)$ are
directly related to the sides of the above effective Dirac UTs. This
point will become clear in section 4.

With the help of Eqs. (15), (18) and (32), some straightforward
calculations lead us to the following approximate expressions in the
case of a normal neutrino mass hierarchy:
\begin{eqnarray}
\Delta^{\prime}_{11} \hspace{-0.15cm} & \simeq & \hspace{-0.15cm}
-\Delta^{}_{31} \left[ \frac{1}{2} \alpha + \frac{1}{2}
\left(1 - |U^{}_{e 3}|^2\right) \beta - \frac{1}{2} \epsilon
- \frac{3}{4} |U^{}_{e 3}|^2 \beta \epsilon - \frac12
|U^{}_{e 3}|^2 \beta^2\right] \; ,
\nonumber \\
\Delta^{\prime}_{21} \hspace{-0.15cm} & \simeq & \hspace{-0.15cm}
\Delta^{}_{31} \left[ \frac{1}{2} \alpha - \frac{1}{2}
\left(1 - |U^{}_{e 3}|^2\right) \beta + \frac{1}{2} \epsilon +
\frac{3}{4} |U^{}_{e 3}|^2 \beta \epsilon + \frac12
|U^{}_{e 3}|^2 \beta^2\right] \; ,
\nonumber \\
\Delta^{\prime}_{22} \hspace{-0.15cm} & \simeq & \hspace{-0.15cm}
\Delta^{}_{31} \left[ \frac{1}{2} \alpha - \frac{1}{2}
\left(1 - |U^{}_{e 3}|^2\right) \beta - \frac{1}{2} \epsilon
- \frac{3}{4} |U^{}_{e 3}|^2 \beta \epsilon + \frac12
|U^{}_{e 3}|^2 \beta^2\right] \; ,
\nonumber \\
\Delta^{\prime}_{31} \hspace{-0.15cm} & \simeq & \hspace{-0.15cm}
\Delta^{}_{31} \left[ 1 - \frac{1}{2} \alpha - \frac{1}{2}
\left(1 - |U^{}_{e 3}|^2\right) \beta + \frac{1}{2} \epsilon
+ \frac{3}{4} |U^{}_{e 3}|^2 \beta \epsilon + \frac12
|U^{}_{e 3}|^2 \beta^2\right] \; ,
\nonumber \\
\Delta^{\prime}_{32} \hspace{-0.15cm} & \simeq & \hspace{-0.15cm}
\Delta^{}_{31} \left[ 1 - \frac{1}{2} \alpha - \frac{1}{2}
\left(1 - |U^{}_{e 3}|^2\right) \beta - \frac{1}{2} \epsilon -
\frac{3}{4} |U^{}_{e 3}|^2 \beta \epsilon + \frac12
|U^{}_{e 3}|^2 \beta^2\right] \; ,
\end{eqnarray}
together with $\Delta^\prime_{13} \simeq -\Delta^{}_{31} \left(1 +
|U^{}_{e 3}|^2 \beta\right)$ and $\Delta^\prime_{33} \simeq
-\Delta^{}_{31} |U^{}_{e 3}|^2 \beta$. When an inverted neutrino
mass ordering is concerned, the corresponding expressions of
$\Delta^\prime_{ij}$ can simply be obtained from Eq. (34) with the
replacement $\epsilon \to -\epsilon$. Given Eqs. (18), (19) and
(34), it is easy to make analytical approximations to the sides of
three Dirac UTs in Eqs. (29)---(31). In the case of a normal mass
hierarchy, we arrive at
\begin{eqnarray}
\widetilde{\triangle}_e : ~ \left\{
\begin{array}{lcl}
\widetilde{U}^{}_{\mu 1} \widetilde{U}^*_{\tau 1} \hspace{-0.15cm} &
\simeq & \hspace{-0.15cm} \displaystyle \frac{\alpha}{\epsilon} \
U^{}_{\mu 1} U^*_{\tau 1} - \frac{1}{2} \left(1 - \frac{\alpha
-\beta}{\epsilon}\right) U^{}_{\mu 3} U^*_{\tau 3} \; ,
\\ \vspace{-0.4cm} \\
\widetilde{U}^{}_{\mu 2} \widetilde{U}^*_{\tau 2} \hspace{-0.15cm} &
\simeq & \hspace{-0.15cm} \displaystyle \frac{\alpha}{\epsilon} \
U^{}_{\mu 2} U^*_{\tau 2} - \frac{1}{2} \left(1 - \frac{\alpha
+\beta}{\epsilon}\right) U^{}_{\mu 3} U^*_{\tau 3} \; ,
\\ \vspace{-0.4cm} \\
\widetilde{U}^{}_{\mu 3} \widetilde{U}^*_{\tau 3} \hspace{-0.15cm} &
\simeq & \hspace{-0.15cm} \displaystyle \left(1 - 2 \beta
\sin^2\theta^{}_{13} + \alpha\beta \sin^2 \theta^{}_{12}
\right) U^{}_{\mu 3} U^*_{\tau 3} + \alpha\beta^{} U^{}_{\mu 1}
U^*_{\tau 1} \; ;
\end{array} \right.
\end{eqnarray}
and
\begin{eqnarray}
\widetilde{\triangle}_\mu : ~ \left\{
\begin{array}{lcl}
\widetilde{U}^{}_{\tau 1} \widetilde{U}^*_{e 1} \hspace{-0.15cm} &
\simeq & \hspace{-0.15cm} \displaystyle \frac{\alpha}{\epsilon} \
U^{}_{\tau 1} U^*_{e 1} - \frac{1}{2} \left(1 + \beta - \frac{\alpha
+\beta}{\epsilon}\right) U^{}_{\tau 3} U^*_{e 3} \; ,
\\ \vspace{-0.4cm} \\
\widetilde{U}^{}_{\tau 2} \widetilde{U}^*_{e 2} \hspace{-0.15cm} &
\simeq & \hspace{-0.15cm} \displaystyle \frac{\alpha}{\epsilon} \
U^{}_{\tau 2} U^*_{e 2} - \frac{1}{2} \left(1 + \beta - \frac{\alpha
-\beta}{\epsilon}\right) U^{}_{\tau 3} U^*_{e 3} \; ,
\\ \vspace{-0.4cm} \\
\widetilde{U}^{}_{\tau 3} \widetilde{U}^*_{e 3} \hspace{-0.15cm} &
\simeq & \hspace{-0.15cm} \displaystyle \left(1 + \beta\right)
U^{}_{\tau 3} U^*_{e 3} - \alpha\beta \sin^2\theta^{}_{13}
U^{}_{\tau 1} U^*_{e 1} \; ; \hspace{2.8cm}
\end{array} \right.
\end{eqnarray}
and
\begin{eqnarray}
\widetilde{\triangle}_\tau : ~ \left\{
\begin{array}{lcl}
\widetilde{U}^{}_{e 1} \widetilde{U}^*_{\mu 1} \hspace{-0.15cm} &
\simeq & \hspace{-0.15cm} \displaystyle \frac{\alpha}{\epsilon} \
U^{}_{e 1} U^*_{\mu 1} - \frac{1}{2} \left(1 + \beta - \frac{\alpha
+\beta}{\epsilon}\right) U^{}_{e 3} U^*_{\mu 3} \; ,
\\ \vspace{-0.4cm} \\
\widetilde{U}^{}_{e 2} \widetilde{U}^*_{\mu 2} \hspace{-0.15cm} &
\simeq & \hspace{-0.15cm} \displaystyle \frac{\alpha}{\epsilon} \
U^{}_{e 2} U^*_{\mu 2} - \frac{1}{2} \left(1 + \beta - \frac{\alpha
-\beta}{\epsilon}\right) U^{}_{e 3} U^*_{\mu 3} \; ,
\\ \vspace{-0.4cm} \\
\widetilde{U}^{}_{e 3} \widetilde{U}^*_{\mu 3} \hspace{-0.15cm} &
\simeq & \hspace{-0.15cm} \displaystyle \left(1 + \beta\right)
U^{}_{e 3} U^*_{\mu 3} - \alpha\beta \sin^2\theta^{}_{13} U^{}_{e
1} U^*_{\mu 1} \; . \hspace{2.8cm}
\end{array} \right.
\end{eqnarray}
One may check that the unitarity of each of the effective triangles
holds up to the ${\cal O}(\alpha\beta)$, ${\cal
O}(\sin^2\theta^{}_{13} \beta)$ or higher-order corrections in the
above approximations. At this precision level the deviation of
$\widetilde{\triangle}^{}_\mu$ from $\triangle^{}_\mu$ and the
departure of $\widetilde{\triangle}^{}_\tau$ from
$\triangle^{}_\tau$ are exactly the same, reflecting a kind of
$\mu$-$\tau$ flavor symmetry between these two effective UTs
\cite{Zhao}. Some further discussions about our approximate
analytical results are in order.

(a) The $\widetilde{U}^{}_{\alpha 3} \widetilde{U}^*_{\beta 3}$ side
of the UT $\widetilde{\triangle}^{}_\gamma$, where the subscripts
$\alpha$, $\beta$ and $\gamma$ run cyclically over $e$, $\mu$ and
$\tau$, is least sensitive to terrestrial matter effects. The reason
is simply that $\widetilde{U}^{}_{e 3}$, $\widetilde{U}^{}_{\mu 3}$
and $\widetilde{U}^{}_{\tau 3}$ only depend on the effective flavor
mixing angles $\widetilde{\theta}^{}_{13} \simeq \theta^{}_{13}$ and
$\widetilde{\theta}^{}_{23} \simeq \theta^{}_{23}$, which are almost
insensitive to the matter-induced corrections when the neutrino beam
energy $E$ is low. Hence $\widetilde{U}^{}_{\alpha 3}
\widetilde{U}^*_{\beta 3} \simeq U^{}_{\alpha 3} U^*_{\beta 3}$ is a
reasonably good approximation for $\widetilde{\triangle}^{}_\gamma$
and $\triangle^{}_\gamma$ in the low-energy region. In other words,
the size and orientation of this side are essentially stable against
terrestrial matter effects, and thus the deviation of
$\widetilde{\triangle}^{}_\gamma$ from $\triangle^{}_\gamma$ is
mainly attributed to the changes of the other two sides.

(b) The $\widetilde{U}^{}_{\alpha 1} \widetilde{U}^*_{\beta 1}$ (or
$\widetilde{U}^{}_{\alpha 2} \widetilde{U}^*_{\beta 2}$) side of
$\widetilde{\triangle}^{}_\gamma$ consists of the corresponding
$U^{}_{\alpha 1} U^*_{\beta 1}$ (or $U^{}_{\alpha 2} U^*_{\beta 2}$)
side of $\triangle^{}_\gamma$ multiplied by a universal factor
$\alpha/\epsilon$ and the $U^{}_{\alpha 3} U^*_{\beta 3}$ side of
$\triangle^{}_\gamma$ multiplied by another factor. Because of
\begin{eqnarray}
\frac{|\alpha|}{\epsilon} \simeq \frac{1}{\sqrt{\displaystyle 1 -
2\cos 2\theta^{}_{12} \cos^2\theta^{}_{13} \frac{A}{\Delta^{}
_{21}} + \cos^4 \theta^{}_{13} \left(\frac{A}{\Delta^{}_{21}}
\right)^2}} \; , \hspace{1cm}
\end{eqnarray}
it becomes clear that this factor approaches $1$ for $A \to 0$ and
approximates to $1/\left(2 \sin\theta^{}_{12}\right) \simeq 0.91$
when $A \simeq \Delta^{}_{21}$ holds (i.e., $\alpha \simeq \beta$
with $E\simeq 0.33$ GeV) for a {\it neutrino} beam or to $1/\left(2
\cos\theta^{}_{12}\right) \simeq 0.60$ when $A \simeq
\Delta^{}_{21}$ holds for an {\it antineutrino} beam. In comparison,
the term proportional to $U^{}_{\alpha 3} U^*_{\beta 3}$ can change
the orientation of the $\widetilde{U}^{}_{\alpha 1}
\widetilde{U}^*_{\beta 1}$ (or $\widetilde{U}^{}_{\alpha 2}
\widetilde{U}^*_{\beta 2}$) side, and its factors $\left[1 -
\left(\alpha \mp \beta\right)/\epsilon \right]/2$ may appreciably
deviate from zero even though $E$ is small. The deformation of the
UT $\triangle^{}_\gamma$ is therefore understandable.

(c) Note that the approximate analytical results in Eqs. (35)---(38)
are valid for the normal neutrino mass hierarchy. As for the
inverted neutrino mass hierarchy with $\alpha <0$, the nine sides of
the three effective Dirac UTs can be directly read off from Eqs.
(35)---(37) with the replacement $\epsilon \to -\epsilon$. In this
case one may similarly discuss the deformation of each UT in the
low-energy region for either a neutrino beam or an antineutrino
beam.

To be more explicit, let us look at the unique peak $\widetilde{\cal
J}^{}_*/{\cal J} \simeq |\alpha|/\epsilon^{}_* \simeq 1/\sin
2\theta^{}_{12}$ at the resonance point $\beta^{}_* \simeq \alpha
\cos 2\theta^{}_{12}$ in the leading-order approximation, as already
discussed below Eq. (22). In this special but interesting case the
nine sides of the effective Dirac UTs can be simply expressed as
follows
\footnote{In this case the matter-induced corrections to the three
Dirac UTs are not very significant due to the smallness of $E^{}_*$,
but the corresponding analytical approximations are simple and
instructive because they only involve a single known parameter
$\theta^{}_{12}$ at the leading-order level.}:
\begin{eqnarray}
\widetilde{\triangle}^{}_e : ~ \left\{
\begin{array}{lcl}
\widetilde{U}^{}_{\mu 1} \widetilde{U}^*_{\tau 1} \hspace{-0.15cm}
& \simeq & \hspace{-0.15cm} \displaystyle \frac{1}{\sin
2\theta^{}_{12}} \ U^{}_{\mu 1} U^*_{\tau 1} - \frac{1 -
\tan\theta^{}_{12}}{2} \ U^{}_{\mu 3} U^*_{\tau 3} \simeq 1.09 \
U^{}_{\mu 1} U^*_{\tau 1} - 0.17 \ U^{}_{\mu 3} U^*_{\tau 3} \; ,
\\ \vspace{-0.4cm} \\
\widetilde{U}^{}_{\mu 2} \widetilde{U}^*_{\tau 2} \hspace{-0.15cm}
& \simeq & \hspace{-0.15cm} \displaystyle \frac{1}{\sin
2\theta^{}_{12}} \ U^{}_{\mu 2} U^*_{\tau 2} - \frac{1 -
\cot\theta^{}_{12}}{2} \ U^{}_{\mu 3} U^*_{\tau 3} \simeq 1.09 \
U^{}_{\mu 2} U^*_{\tau 2} + 0.26 \ U^{}_{\mu 3} U^*_{\tau 3} \; ,
\\ \vspace{-0.4cm} \\
\widetilde{U}^{}_{\mu 3} \widetilde{U}^*_{\tau 3} \hspace{-0.15cm}
&\simeq & \hspace{-0.15cm} U^{}_{\mu 3} U^*_{\tau 3} \; ;
\end{array} \right.
\end{eqnarray}
and
\begin{eqnarray}
\widetilde{\triangle}^{}_\mu : ~ \left\{
\begin{array}{lcl}
\widetilde{U}^{}_{\tau 1} \widetilde{U}^*_{e 1} \hspace{-0.15cm} &
\simeq & \hspace{-0.15cm} \displaystyle \frac{1}{\sin
2\theta^{}_{12}} \ U^{}_{\tau 1} U^*_{e 1} - \frac{1 -
\cot\theta^{}_{12}}{2} \ U^{}_{\tau 3} U^*_{e 3} \simeq 1.09 \
U^{}_{\tau 1} U^*_{e 1} + 0.26 \ U^{}_{\tau 3} U^*_{e 3} \; ,
\\ \vspace{-0.4cm} \\
\widetilde{U}^{}_{\tau 2} \widetilde{U}^*_{e 2} \hspace{-0.15cm} &
\simeq & \hspace{-0.15cm} \displaystyle \frac{1}{\sin
2\theta^{}_{12}} \ U^{}_{\tau 2} U^*_{e 2} - \frac{1 -
\tan\theta^{}_{12}}{2} \ U^{}_{\tau 3} U^*_{e 3} \simeq 1.09 \
U^{}_{\tau 2} U^*_{e 2} - 0.17 \ U^{}_{\tau 3} U^*_{e 3} \; ,
\\ \vspace{-0.4cm} \\
\widetilde{U}^{}_{\tau 3} \widetilde{U}^*_{e 3} \hspace{-0.15cm} &
\simeq & \hspace{-0.15cm} U^{}_{\tau 3} U^*_{e 3} \; ;
\end{array} \right.
\end{eqnarray}
and
\begin{eqnarray}
\widetilde{\triangle}^{}_\tau : ~ \left\{
\begin{array}{lcl}
\widetilde{U}^{}_{e 1} \widetilde{U}^*_{\mu 1} \hspace{-0.15cm} &
\simeq & \hspace{-0.15cm} \displaystyle \frac{1}{\sin
2\theta^{}_{12}} \ U^{}_{e 1} U^*_{\mu 1} - \frac{1 -
\cot\theta^{}_{12}}{2} \ U^{}_{e 3} U^*_{\mu 3} \simeq 1.09 \
U^{}_{e 1} U^*_{\mu 1} + 0.26 \ U^{}_{e 3} U^*_{\mu 3} \; ,
\\ \vspace{-0.4cm} \\
\widetilde{U}^{}_{e 2} \widetilde{U}^*_{\mu 2} \hspace{-0.15cm} &
\simeq & \hspace{-0.15cm} \displaystyle \frac{1}{\sin
2\theta^{}_{12}} \ U^{}_{e 2} U^*_{\mu 2} - \frac{1 -
\tan\theta^{}_{12}}{2} \ U^{}_{e 3} U^*_{\mu 3} \simeq 1.09 \
U^{}_{e 2} U^*_{\mu 2} - 0.17 \ U^{}_{e 3} U^*_{\mu 3} \; ,
\\ \vspace{-0.4cm} \\
\widetilde{U}^{}_{e 3} \widetilde{U}^*_{\mu 3} \hspace{-0.15cm} &
\simeq & \hspace{-0.15cm} U^{}_{e 3} U^*_{\mu 3} \; ,
\end{array} \right.
\end{eqnarray}
where $\theta^{}_{12} \simeq 33.48^\circ$ has been taken as a
typical input value to illustrate the deviation of each effective UT
from its fundamental counterpart in vacuum. In particular, the
enhancement of $\widetilde{\cal J}$ and the deformation of each
triangle become quite transparent. For example, Eq. (41) leads us to
the approximate relationship
\begin{eqnarray}
\widetilde{\cal J}^{}_* = {\rm Im}\left(\widetilde{U}^{}_{e 2}
\widetilde{U}^{}_{\mu 3} \widetilde{U}^*_{e 3} \widetilde{U}^*_{\mu 2}
\right) \simeq \frac{1}{\sin 2\theta^{}_{12}}
{\rm Im}\left(U^{}_{e2} U^{}_{\mu 3} U^*_{ e3} U^*_{\mu 2}\right)
= \frac{1}{\sin 2\theta^{}_{12}} {\cal J} \;
\end{eqnarray}
at the resonance point $\beta^{}_* \simeq \alpha \cos
2\theta^{}_{12}$ under discussion. In Fig. 4 we plot the three Dirac
UTs in the complex plane by inputting the best-fit values of six
neutrino oscillation parameters and taking the resonant beam energy
$E^{}_* \simeq 0.140$ GeV (or $0.123$ GeV) for the normal (or
inverted) neutrino mass ordering, corresponding to the peak of
$\widetilde{\cal J}/{\cal J}$ shown in Fig. 2. Now the deformation
of each UT becomes more intuitive, although the terrestrial matter
effects in such a low-energy case are not very significant. Two
comments are in order.
\begin{figure*}
\vspace{0.2cm}
\centerline{\includegraphics[width=14.5cm]{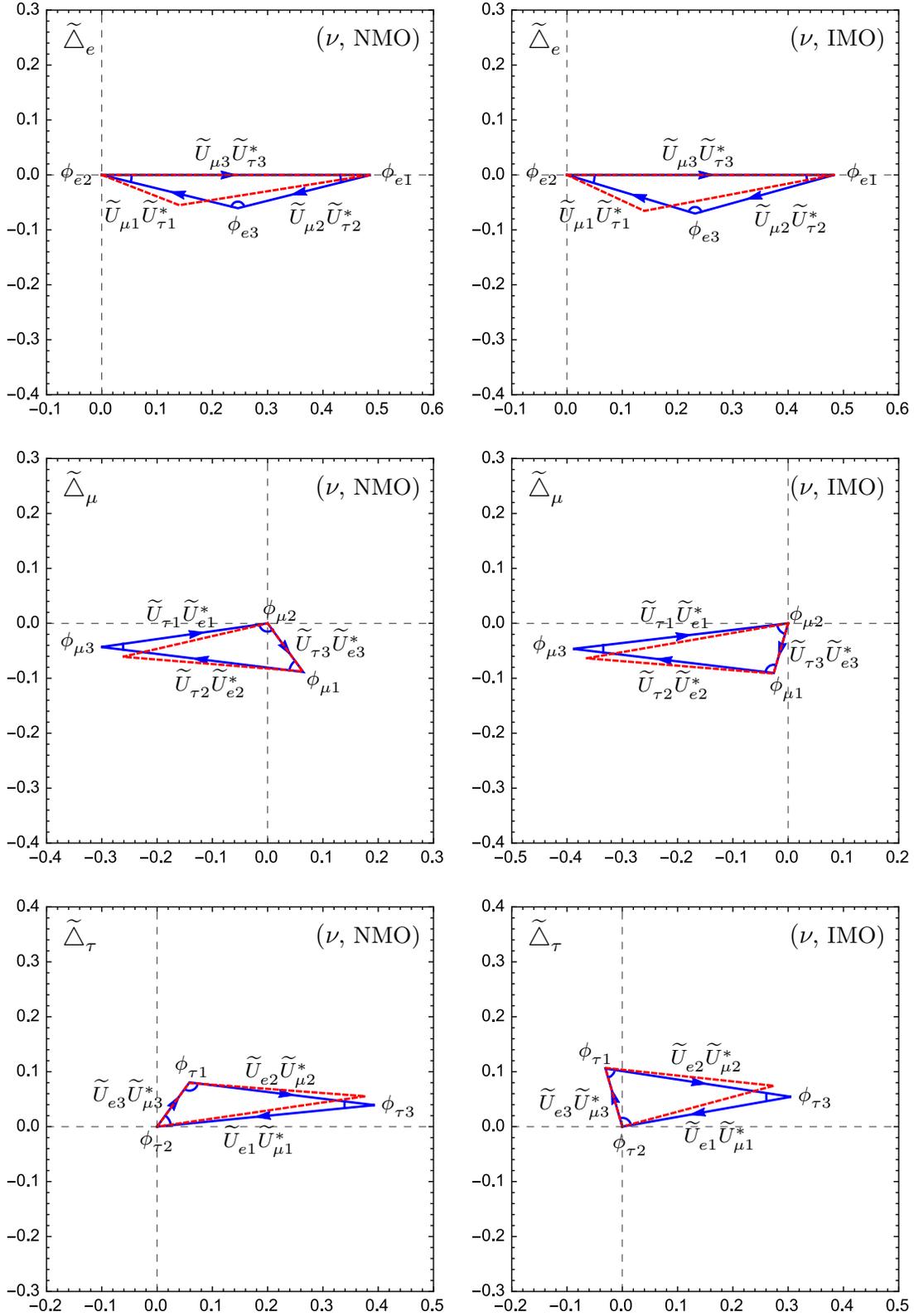}}
\caption{The matter-deformed Dirac UTs (blue and solid) as compared
with their counterparts in vacuum (red and dashed), where the
best-fit values of six neutrino oscillation parameters \cite{GG}
have been input and the resonant beam energy $E^{}_* \simeq 0.140$
GeV (or $0.123$ GeV) has been taken for the normal (or inverted)
neutrino mass ordering, corresponding to the peak of
$\widetilde{\cal J}/{\cal J}$ shown in Fig. 2.}
\end{figure*}

(1) Fig. 4 is a reflection of the real shapes of the fundamental and
effective UTs based on the best-fit results of current neutrino
oscillation data. The configurations of $\triangle^{}_\mu$ (or
$\triangle^{}_\tau$) in the cases of normal and inverted mass
neutrino hierarchies look quite different, simply because one of its
three sides is proportional to $U^*_{e 3} = s^{}_{13} e^{{\rm
i}\delta}$ (or its complex conjugate) but the best-fit value of the
CP-violating phase $\delta$ lies in two different quadrants in these
two cases, as one can see in Table 1. In comparison, the
configuration of $\triangle^{}_e$ is not so sensitive to the
best-fit values of $\delta$ in the cases of normal and inverted mass
hierarchies since its three sides do not directly depend on $U^{}_{e
3}$ or equivalently $s^{}_{13} e^{-{\rm i}\delta}$. As for the three
effective Dirac UTs in matter, the same observations are true.

(2) Although the matter-induced corrections to the three fundamental
UTs are not very significant, one can see a clear change in the
orientations of two sides of each triangle at the resonance energy
$E^{}_*$. Our numerical results in Fig. 4 confirm the observations
based on the analytical approximations made below Eq. (37), implying
that we have fully understood the matter-corrected behaviors of
leptonic CP and T violation in the low-energy region.

To be more realistic, Figs. 5 and 6 show the matter-corrected Dirac
UTs corresponding to the realistic accelerator-based T2K \cite{T2K}
and NO$\nu$A \cite{NOVA} experiments which have the typical neutrino
beam energies $0.6$ GeV and $2$ GeV, respectively. In plotting these
two figures we have input the best-fit values of six oscillation
parameters and considered both the neutrino and antineutrino beams.
Note that a description of {\it antineutrino} oscillations in matter
actually involves $U^*$ and $-A$, but here we plot the relevant
effective UTs defined in Eq. (28) with $-A$ instead of their complex
conjugate counterparts for an {\it antineutrino} beam so as to make
a direct comparison between the same set of triangles in the
neutrino (Fig. 5 with $A$) and antineutrino (Fig. 6 with $-A$) cases
\footnote{This point can be easily understood as follows. For
example, $\widetilde{\triangle}^{}_e$ is defined by the
orthogonality relation $\widetilde{U}^{}_{\mu 1}
\widetilde{U}^*_{\tau 1} + \widetilde{U}^{}_{\mu 2}
\widetilde{U}^*_{\tau 2} + \widetilde{U}^{}_{\mu 3}
\widetilde{U}^*_{\tau 3} = 0$ for a neutrino beam depending on $U$
and $A$. As for an antineutrino beam depending on $U^*$ and $-A$,
the corresponding effective triangle is described by
$\widetilde{U}^{*}_{\mu 1} \widetilde{U}^{}_{\tau 1} +
\widetilde{U}^{*}_{\mu 2} \widetilde{U}^{}_{\tau 2} +
\widetilde{U}^{*}_{\mu 3} \widetilde{U}^{}_{\tau 3} = 0$. What we
have done in plotting Fig. 6 is simply to make a complex conjugation
of this orthogonality relation, such that
$\widetilde{\triangle}^{}_e$ as a function of $A$ in Fig. 5 and
$\widetilde{\triangle}^{}_e$ as a function of $-A$ in Fig. 6 can be
directly compared.}.
Some comments and discussions are in order.
\begin{figure*}
\vspace{-0.3cm}
\centerline{\includegraphics[width=15cm]{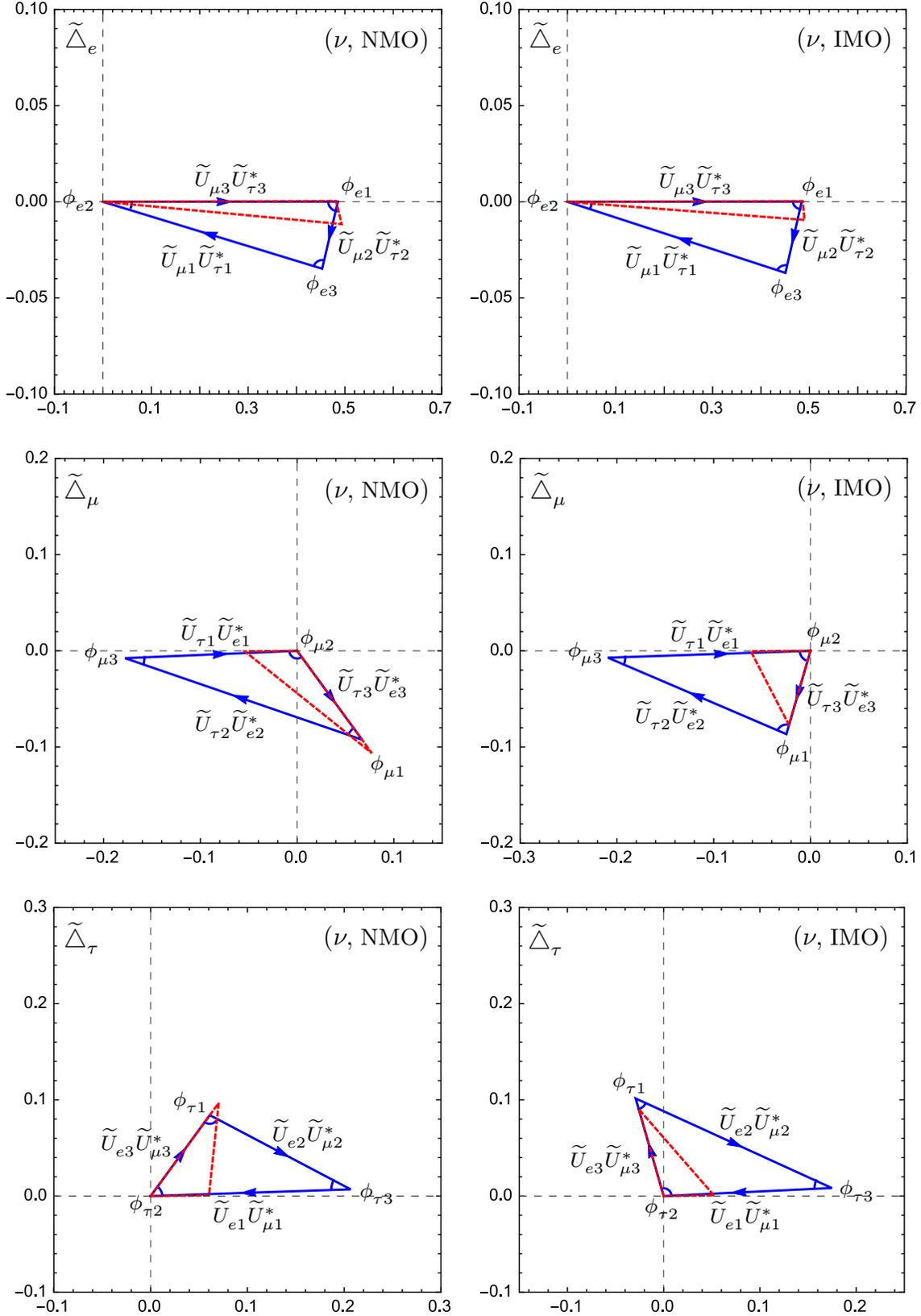}} \caption{The
real shapes of three matter-corrected Dirac UTs corresponding to the
T2K (blue and solid, $E\simeq 0.6$ GeV) and NO$\nu$A (red and
dashed, $E\simeq 2$ GeV) {\it neutrino} oscillation experiments,
where the best-fit values of six oscillation parameters \cite{GG}
have been input.}
\end{figure*}
\begin{figure*}
\vspace{-0.3cm}
\centerline{\includegraphics[width=15cm]{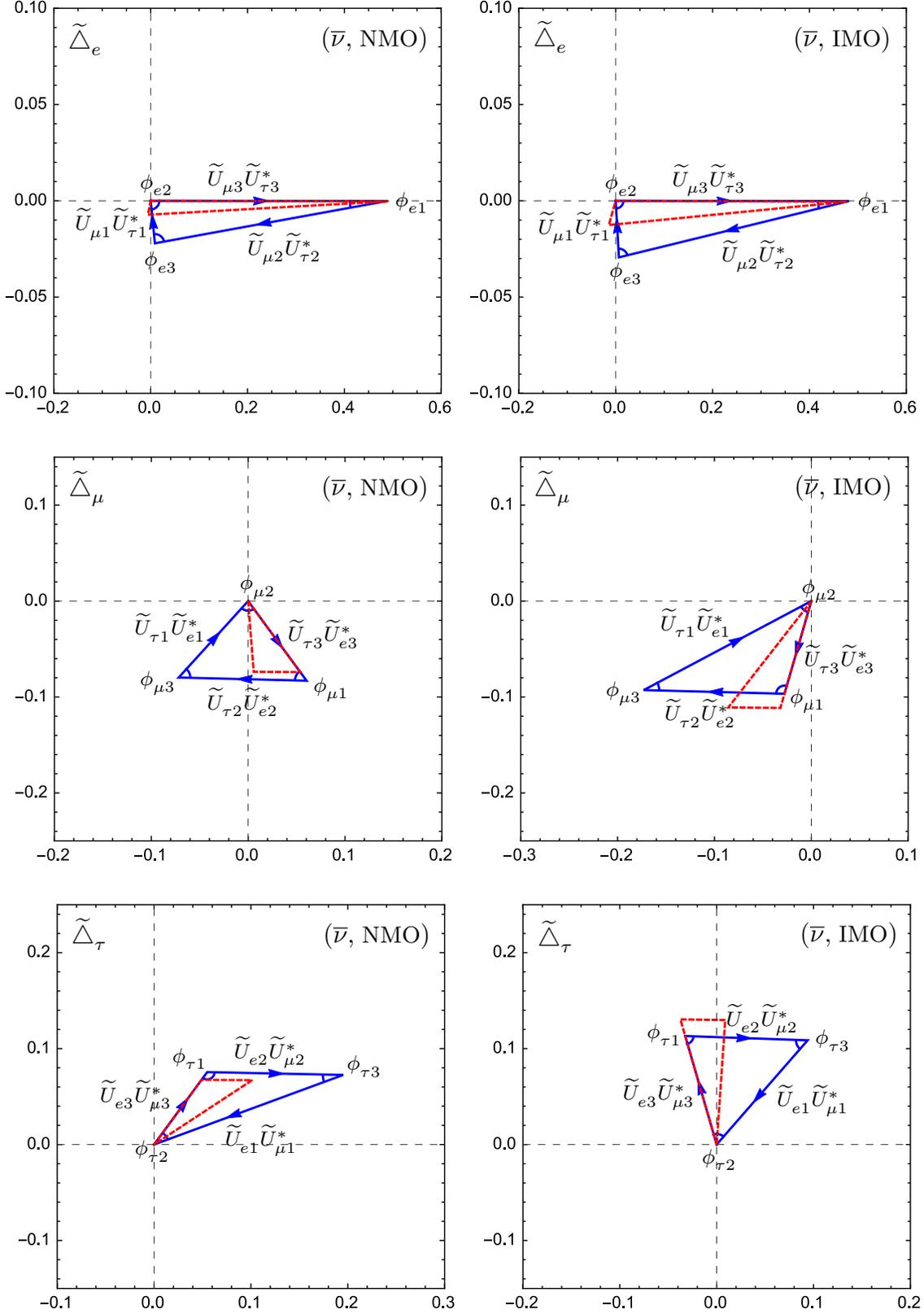}} \caption{The
real shapes of three matter-corrected Dirac UTs corresponding to the
T2K (blue and solid, $E\simeq 0.6$ GeV) and NO$\nu$A (red and
dashed, $E\simeq 2$ GeV) {\it antineutrino} oscillation experiments,
where the same inputs as those in Fig. 5 have been used.}
\end{figure*}

(1) In both the T2K and NO$\nu$A cases, the ratios of
$\widetilde{\cal J}$ to ${\cal J}$ are strongly suppressed, as one
can expect from Fig. 1. To be specific, we obtain
\begin{eqnarray}
{\rm T2K} ~(E \simeq 0.6 ~{\rm GeV}): &&
\frac{\widetilde{\cal J}}{\cal J} \simeq \left\{
\begin{array}{l}
0.633 \hspace{0.6cm} (\nu ~ {\rm beam}, \Delta^{}_{31} >0) \; , \\
0.568 \hspace{0.6cm} (\nu ~ {\rm beam}, \Delta^{}_{31} <0) \; , \\
0.402 \hspace{0.6cm} (\overline{\nu} ~ {\rm beam}, \Delta^{}_{31}
>0) \; , \\ 0.448 \hspace{0.6cm} (\overline{\nu} ~ {\rm beam},
\Delta^{}_{31} <0) \; ;
\end{array} \right.
\nonumber \\
{\rm NO}\nu{\rm A} ~(E \simeq 2 ~{\rm GeV}): &&
\frac{\widetilde{\cal J}}{\cal J} \simeq \left\{
\begin{array}{l}
0.216 \hspace{0.6cm} (\nu ~ {\rm beam}, \Delta^{}_{31} >0) \; , \\
0.150 \hspace{0.6cm} (\nu ~ {\rm beam}, \Delta^{}_{31} <0) \; , \\
0.132 \hspace{0.6cm} (\overline{\nu} ~ {\rm beam}, \Delta^{}_{31}
>0) \; , \\ 0.190 \hspace{0.6cm} (\overline{\nu} ~ {\rm beam},
\Delta^{}_{31} <0) \; , \hspace{0.5cm}
\end{array} \right.
\end{eqnarray}
where ${\cal J} \simeq -0.0268$ (normal hierarchy) or $-0.0316$
(inverted hierarchy), calculated by inputting the best-fit values of
$\theta^{}_{12}$, $\theta^{}_{13}$, $\theta^{}_{23}$ and $\delta$ as
listed in Table 1. Hence the areas of the UTs in Figs. 5 and 6 are
much smaller than those in Fig. 4, where $\widetilde{\cal
J}^{}_*/{\cal J} \simeq 1.10$ (normal hierarchy) or $1.07$ (inverted
hierarchy) for a neutrino beam with the same inputs.

(2) The $\widetilde{U}^{}_{\alpha 3} \widetilde{U}^*_{\beta 3}$ side
of $\widetilde{\triangle}^{}_\gamma$, where the subscripts $\alpha$,
$\beta$ and $\gamma$ run cyclically over $e$, $\mu$ and $\tau$,
remains least sensitive to terrestrial matter effects as compared
with the other two sides when the neutrino (or antineutrino) beam
energy goes up. In either Fig. 5 or Fig. 6, one may see the
difference between the shapes or orientations of the same UT in the
normal and inverted neutrino mass hierarchy cases. Such a difference
mainly originates from the fact that the best-fit values of $\delta$
lie in two different quadrants in these two cases. When comparing
one UT with respect to a neutrino beam with its counterpart with
respect to an antineutrino beam (i.e., one UT in Fig. 5 and its
counterpart in Fig. 6), we find that the changes associated with
each triangle's shapes and orientations corresponding to $A
\leftrightarrow -A$ are appreciable and even significant when $E$
increases --- this effect is just the {\it fake} CP-violating
asymmetry induced by terrestrial matter between $\nu^{}_\alpha \to
\nu^{}_\beta$ and $\overline{\nu}^{}_\alpha \to
\overline{\nu}^{}_\beta$ oscillations. The reason for this
``asymmetry" is simply that the ordinary matter background is not
symmetric under the CP transformation.

(3) The nine inner angles of the three effective Dirac UTs in matter
can be defined as $\phi^{}_{\alpha i} \equiv \arg\left[-
\left(\widetilde{U}^{}_{\beta j} \widetilde{U}^*_{\gamma
j}\right)/\left(\widetilde{U}^{}_{\beta k} \widetilde{U}^*_{\gamma
k}\right)\right]$ \cite{Luo}, where the Greek and Latin subscripts
keep their cyclic running over $(e, \mu, \tau)$ and $(1, 2, 3)$,
respectively. Taking the T2K and NO$\nu$A experiments for example,
we calculate these inner angles and list the numerical results in
Table 2, where the best-fit values of six neutrino oscillation
parameters shown in Table 1 have been input. It is obvious that
terrestrial matter effects may significantly change the inner angles
of the UTs, and therefore change their configurations and
orientations.
\begin{table}[t]
\centering \caption{A numerical illustration of terrestrial matter
effects on the inner angles of the Dirac UTs in the T2K (with $E
\simeq 0.6$ GeV) and NO$\nu$A (with $E\simeq 2$ GeV) experiments,
where the best-fit values of six neutrino oscillation parameters
\cite{GG} have been typically input.} \vspace{0.4cm}
\begin{tabular}[b]{c|lll|lll}
\hline & \multicolumn{3}{c|}{Normal mass ordering (NMO)} &
\multicolumn{3}{c}{Inverted mass ordering (IMO)} \\
&&&&&&\\ \vspace{-1.15cm} \\
& $\widetilde U=U$ & $E\simeq 0.6~\rm{GeV}$ & $E\simeq 2~\rm{GeV}$ &
$ \widetilde U=U$ & $E\simeq 0.6~\rm{GeV}$ & $E\simeq 2~\rm{GeV}$ \\
\hline
$\phi_{e1}^{}$ & $9.0^{\circ}$ & $ \left\{
\begin{array}{r} 47.0^{\circ} ~(\nu) \\ 2.6^{\circ} ~(\overline{\nu})
\end{array} \right.$ & $\left\{ \begin{array}{r}
134.8^{\circ} ~(\nu) \\ 0.8^{\circ} ~(\overline{\nu}) \end{array}
\right.$ & $10.8^{\circ}$ & $\left\{ \begin{array}{r} 47.8^{\circ}
~(\nu) \\ 3.5^{\circ} ~(\overline{\nu}) \end{array} \right.$ &
$\left\{ \begin{array}{r} 111.7^ {\circ} ~(\nu) \\ 1.5^{\circ}
~(\overline{\nu}) \end{array} \right.$ \\ \hline
$\phi_{e2}^{}$ & $21.4^{\circ}$ & $\left\{ \begin{array}{r}
4.4^{\circ} ~(\nu) \\ 68.4^{\circ} ~(\overline{\nu}) \end{array}
\right.$
& $\left\{ \begin{array}{r} 1.4^{\circ} ~(\nu) \\
128.0^{\circ} ~(\overline{\nu}) \end{array} \right.$ &
$25.1^{\circ}$ & $\left\{ \begin{array}{r} 4.7^{\circ} ~(\nu) \\
77.4^{\circ} ~(\overline{\nu}) \end{array} \right.$ & $\left\{
\begin{array}{r} 1.2^{\circ} ~(\nu) \\ 138.3^{\circ}
~(\overline{\nu}) \end{array} \right.$ \\ \hline
$\phi_{e3}^{}$ & $149.6^{\circ}$ & $\left\{ \begin{array}{r}
128.6^{\circ} ~(\nu) \\ 109.0^{\circ} ~(\overline{\nu})
\end{array} \right.$ & $\left\{ \begin{array}{r} 43.8^{\circ}
~(\nu)\\
51.2^{\circ} ~(\overline{\nu}) \end{array} \right.$ &
$144.1^{\circ}$
& $\left\{ \begin{array}{r} 127.5^{\circ} ~(\nu) \\ 99.1^{\circ}
~(\overline{\nu})
\end{array} \right.$ & $\left\{ \begin{array}{r} 67.1^{\circ}
~(\nu)\\
40.2^{\circ} ~(\overline{\nu}) \end{array} \right.$ \\ \hline
$\phi_{\mu1}^{}$ & $49.3^{\circ}$ & $\left\{ \begin{array}{r}
34.9^{\circ} ~(\nu) \\ 52.6^{\circ} ~(\overline{\nu}) \end{array}
\right.$ & $\left\{ \begin{array}{r} 15.2^{\circ} ~(\nu) \\
53.6^{\circ} ~(\overline{\nu})
\end{array} \right.$ & $101.3^{\circ}$ & $\left\{ \begin{array}{r}
82.7^{\circ} ~(\nu) \\ 104.5^{\circ} ~(\overline{\nu}) \end{array}
\right.$ & $\left\{ \begin{array}{r} 43.2^{\circ} ~(\nu) \\
105.3^{\circ} ~( \overline{\nu}) \end{array} \right.$ \\ \hline
$\phi_{\mu2}^{}$ & $112.9^{\circ}$ & $\left\{ \begin{array}{r}
123.5^{\circ} ~(\nu) \\ 78.0^{\circ} ~(\overline{\nu}) \end{array}
\right.$ & $\left\{ \begin{array}{r} 125.2^{\circ} ~(\nu) \\
31.5^{\circ} ~(\overline {\nu}) \end{array} \right.$ &
$64.1^{\circ}$ & $\left\{
\begin{array}{r} 72.0^{\circ} ~(\nu) \\ 45.8^{\circ}
~(\overline{\nu}) \end{array} \right.$ & $\left\{ \begin{array}{r}
73.5^{\circ} ~(\nu)
\\
21.8^{\circ} ~(\overline {\nu}) \end{array} \right.$ \\ \hline
$\phi_{\mu3}^{}$ & $17.8^{\circ}$ & $\left\{ \begin{array}{r}
21.6^{\circ} ~(\nu) \\ 49.4^{\circ} ~(\overline{\nu}) \end{array}
\right.$ & $\left\{ \begin{array}{r} 39.6^{\circ} ~(\nu) \\
94.9^{\circ} ~ (\overline{\nu}) \end{array} \right.$ &
$14.6^{\circ}$ & $\left\{ \begin{array}{r} 25.3^{\circ} ~(\nu) \\
29.7^{\circ} ~(\overline{\nu}) \end{array} \right.$
& $\left\{ \begin{array}{r} 63.3^{\circ} ~(\nu) \\
52.9^{\circ} ~(\overline{\nu}) \end{array} \right.$ \\ \hline
$\phi_{\tau1}^{}$ & $121.7^{\circ}$ & $\left\{ \begin{array}{r}
98.1^{\circ} ~(\nu) \\ 124.8^{\circ} ~(\overline{\nu}) \end{array}
\right.$ & $\left\{ \begin{array}{r} 30.0^{\circ} ~(\nu) \\
125.6^{\circ} ~(\overline{\nu}) \end{array} \right.$ &
$67.9^{\circ}$ & $\left\{ \begin{array}{r} 49.5^{\circ} ~(\nu) \\
72.0^{\circ} ~(\overline{\nu})
\end{array} \right.$ & $\left\{ \begin{array}{r} 25.1^{\circ} ~(\nu)
\\
73.2^{\circ} ~(\overline{\nu}) \end{array} \right.$ \\ \hline
$\phi_{\tau2}^{}$ & $45.6^{\circ}$ & $\left\{ \begin{array}{r}
52.1^{\circ} ~(\nu) \\ 33.6^{\circ} ~(\overline{\nu}) \end{array}
\right.$ & $\left\{ \begin{array}{r} 53.4^{\circ} ~(\nu) \\
20.5^{\circ} ~(\overline {\nu}) \end{array} \right.$ &
$90.8^{\circ}$ & $\left\{
\begin{array}{r} 103.3^{\circ} ~(\nu) \\ 56.8^{\circ}
~(\overline{\nu}) \end{array} \right.$ & $\left\{ \begin{array}{r}
105.3^{\circ} ~(\nu)
\\19.9^{\circ} ~(\overline{\nu}) \end{array} \right.$ \\ \hline
$\phi_{\tau3}^{}$ & $12.7^{\circ}$ & $\left\{ \begin{array}{r}
29.8^{\circ} ~(\nu) \\ 21.6^{\circ} ~(\overline{\nu}) \end{array}
\right.$ & $\left\{ \begin{array}{r} 96.6^{\circ} ~(\nu) \\
33.9^{\circ} ~(\overline {\nu}) \end{array} \right.$ &
$21.3^{\circ}$ & $\left\{ \begin{array}{r} 27.2^{\circ} ~(\nu) \\
51.2^{\circ} ~(\overline{\nu}) \end{array} \right.$ & $\left\{
\begin{array}{r} 49.6^{\circ} ~(\nu) \\ 86.9^{\circ} ~(\overline
{\nu}) \end{array} \right.$ \\ \hline
\end{tabular}
\end{table}

\section{Neutrino oscillations and CP violation}

Now we turn to the possibility of measuring leptonic CP violation in
neutrino oscillations in a low-energy or low-matter-density region.
In practice the matter-corrected sides $|\widetilde{U}^{}_{\alpha i}
\widetilde{U}^*_{\beta i}|$ of three Dirac UTs and their
corresponding Jarlskog parameter $\widetilde{\cal J}$ can be
determined from a variety of long- or medium-baseline neutrino
oscillation experiments \cite{Zhang2}. The probabilities of
$\nu^{}_\alpha \to \nu^{}_\beta$ oscillations in matter are given by
\begin{eqnarray}
\widetilde{P}(\nu^{}_\alpha \to \nu^{}_\beta) =
\delta^{}_{\alpha\beta} - 4 \sum_{i<j} {\rm Re}
\left(\widetilde{U}^{}_{\alpha i} \widetilde{U}^{}_{\beta j}
\widetilde{U}^*_{\alpha j} \widetilde{U}^*_{\beta i} \right) \sin^2
\widetilde{F}^{}_{ji} + 8 \widetilde{\cal J} \sum_\gamma
\epsilon^{}_{\alpha\beta\gamma} \prod^{}_{i<j} \sin
\widetilde{F}^{}_{ji} \; ,
\end{eqnarray}
where $\widetilde{F}^{}_{ji} \equiv \widetilde{\Delta}^{}_{ji}
L/\left(4 E\right)$, and the Greek and Latin subscripts run over
$(e, \mu, \tau)$ and $(1, 2, 3)$, respectively. Given the
algebraic relationship
\begin{eqnarray}
{\rm Re} \left(\widetilde{U}^{}_{\alpha i}
\widetilde{U}^{}_{\beta j} \widetilde{U}^*_{\alpha j}
\widetilde{U}^*_{\beta i} \right) = \frac{1}{2} \left(
|\widetilde{U}^{}_{\alpha k} \widetilde{U}^*_{\beta k}|^2 -
|\widetilde{U}^{}_{\alpha i} \widetilde{U}^*_{\beta i}|^2 -
|\widetilde{U}^{}_{\alpha j} \widetilde{U}^*_{\beta j}|^2 \right)
\end{eqnarray}
with $\alpha \neq \beta$ and $i\neq j\neq k$, one may then express
the {\it appearance} ($\beta \neq \alpha$) probabilities of neutrino
oscillations in terms of the sides of the UTs and $\widetilde{\cal
J}$ as follows:
\begin{eqnarray}
\widetilde{P}(\nu^{}_\alpha \to \nu^{}_\beta) \hspace{-0.15cm} & = &
\hspace{-0.15cm} - 2 \left( |\widetilde{U}^{}_{\alpha 3}
\widetilde{U}^*_{\beta 3}|^2 - |\widetilde{U}^{}_{\alpha 1}
\widetilde{U}^*_{\beta 1}|^2 - |\widetilde{U}^{}_{\alpha 2}
\widetilde{U}^*_{\beta 2}|^2 \right) \sin^2 \widetilde{F}^{}_{21}
\hspace{1cm}
\nonumber \\
&& \hspace{-0.15cm} - 2 \left( |\widetilde{U}^{}_{\alpha 2}
\widetilde{U}^*_{\beta 2}|^2 - |\widetilde{U}^{}_{\alpha 1}
\widetilde{U}^*_{\beta 1}|^2 - |\widetilde{U}^{}_{\alpha 3}
\widetilde{U}^*_{\beta 3}|^2 \right) \sin^2 \widetilde{F}^{}_{31}
\nonumber \\
&& \hspace{-0.15cm} - 2 \left( |\widetilde{U}^{}_{\alpha 1}
\widetilde{U}^*_{\beta 1}|^2 - |\widetilde{U}^{}_{\alpha 2}
\widetilde{U}^*_{\beta 2}|^2 - |\widetilde{U}^{}_{\alpha 3}
\widetilde{U}^*_{\beta 3}|^2 \right) \sin^2 \widetilde{F}^{}_{32}
\nonumber \\
&& \hspace{-0.15cm} + 8 \widetilde{\cal J} \sum_\gamma
\epsilon^{}_{\alpha\beta\gamma} \sin\widetilde{F}^{}_{21}
\sin\widetilde{F}^{}_{31} \sin\widetilde{F}^{}_{32} \; .
\end{eqnarray}
Of our particular interest are the $\nu^{}_\mu \to \nu^{}_e$ and
$\overline{\nu}^{}_\mu \to \overline{\nu}^{}_e$ oscillations to
probe leptonic CP violation. In this case it is the Dirac UT
$\widetilde{\triangle}^{}_\tau$ that fully determines the
oscillation probabilities. Namely,
\begin{eqnarray}
\widetilde{P}(\nu^{}_\mu \to \nu^{}_e) \hspace{-0.15cm} & = &
\hspace{-0.15cm} - 2 \left( |\widetilde{U}^{}_{e 3}
\widetilde{U}^*_{\mu 3}|^2 - |\widetilde{U}^{}_{e 1}
\widetilde{U}^*_{\mu 1}|^2 - |\widetilde{U}^{}_{e 2}
\widetilde{U}^*_{\mu 2}|^2 \right) \sin^2 \widetilde{F}^{}_{21}
\hspace{1cm}
\nonumber \\
&& \hspace{-0.15cm} - 2 \left( |\widetilde{U}^{}_{e 2}
\widetilde{U}^*_{\mu 2}|^2 - |\widetilde{U}^{}_{e 1}
\widetilde{U}^*_{\mu 1}|^2 - |\widetilde{U}^{}_{e 3}
\widetilde{U}^*_{\mu 3}|^2 \right) \sin^2 \widetilde{F}^{}_{31}
\nonumber \\
&& \hspace{-0.15cm} - 2 \left( |\widetilde{U}^{}_{e 1}
\widetilde{U}^*_{\mu 1}|^2 - |\widetilde{U}^{}_{e 2}
\widetilde{U}^*_{\mu 2}|^2 - |\widetilde{U}^{}_{e 3}
\widetilde{U}^*_{\mu 3}|^2 \right) \sin^2 \widetilde{F}^{}_{32}
\nonumber \\
&& \hspace{-0.15cm} - 8 \widetilde{\cal J} \sin\widetilde{F}^{}_{21}
\sin\widetilde{F}^{}_{31} \sin\widetilde{F}^{}_{32} \; ,
\end{eqnarray}
and the corresponding expression of
$\widetilde{P}(\overline{\nu}^{}_\mu \to \overline{\nu}^{}_e)$ can
be directly read off from Eq. (47) with the replacements ${\cal J}
\to -{\cal J}$ and $A \to -A$.

To see an interplay between the fundamental physics and terrestrial
matter effects in the probability of $\nu^{}_\mu \to \nu^{}_e$
oscillations in a more transparent way, let us make an analytical
approximation for the expression of $\widetilde{P}(\nu^{}_\mu \to
\nu^{}_e)$ in Eq. (47), whose CP-conserving part is only related to
the sides of $\widetilde{\triangle}^{}_\tau$. Instead of adopting
Eq. (37), here we start from Eq. (31) and make a higher-order
analytical approximation to ensure a sufficient accuracy associated
with $\widetilde P (\nu_{\mu}^{} \to \nu_e^{})$ itself. That is,
\begin{eqnarray}
\widetilde{U}_{e1}^{}\widetilde{U}_{\mu1}^*\hspace{-0.15cm} & = &
\hspace{-0.15cm}\frac{\alpha}{\epsilon}{U}_{e1}^{}{U} _{\mu1}^* +
\frac{\alpha - \epsilon + \beta \cos^2\theta_{13}^{} - \epsilon
\beta -\alpha \beta\cos 2\theta_{12}^{} + \beta^2
}{2\epsilon}{U}_{e3}^{}{U}_{\mu3}^* \; ,
\nonumber\\
\widetilde{U}_{e2}^{}\widetilde{U}_{\mu2}^*\hspace{-0.15cm} & = &
\hspace{-0.15cm}\frac{\alpha}{\epsilon}{U}_{e2}^{}{U} _{\mu2}^* +
\frac{\alpha - \epsilon - \beta \cos^2\theta_{13}^{} - \epsilon
\beta + \alpha \beta\cos 2\theta_{12}^{} - \beta^2
}{2\epsilon}{U}_{e3}^{}{U}_{\mu3}^* \; ,
\nonumber\\
\widetilde{U}_{e3}^{}\widetilde{U}_{\mu3}^*\hspace{-0.15cm} & = &
\hspace{-0.15cm} \left(1+\beta\right) {U}_{e3}^{}{U}_{\mu3}^* \; .
\end{eqnarray}
With the help of Eqs. (18), (21) and (48), we finally arrive at the
result
\begin{eqnarray}
\widetilde P (\nu_{\mu}^{} \to \nu_e^{}) \hspace{-0.15cm} & \simeq &
\hspace{-0.15cm} \displaystyle \alpha^2 \sin^2 2\theta_{12}^{}
\cos^2\theta_{13}^{} \left(\cos^2\theta_{23}^{} - \sin^2
\theta_{13}^{} \sin^2 \theta_{23}^{}\right) \frac{\sin^2 \left(
\epsilon F_{31}^{} \right)}{\epsilon^2}
\nonumber\\
& & \hspace{-0.2cm} \displaystyle + \frac{1}{2} \left(1 + 2\beta\right)
\sin^22\theta_{13}^{} \sin^2\theta_{23}^{} \left[ 1-\cos \left(F^{}_*
- \beta F^{}_{31}\right) \cos \left( \epsilon F_{31}^{} \right)
\right] \nonumber\\
& & \hspace{-0.2cm} \displaystyle + \frac{1}{2} \left(1 + 2\beta\right)
\left(\alpha\cos 2\theta_{12}^{} - \beta \cos^2\theta_{13}^{}\right)
\sin^2 2\theta_{13}^{} \sin^2\theta_{23}^{} \sin \left(F^{}_* -
\beta F_{31}^{}\right) \frac{\sin \left(\epsilon F_{31}^{}
\right)}{\epsilon}  \nonumber\\
& & \hspace{-0.2cm} \displaystyle + 4 {\cal J}\alpha \left(1 + \beta\right)
\left(\alpha \cos2 \theta_{12} - \beta \cos^2 \theta_{13}^{}\right)
\cot\delta\frac{\sin^2 \left(\epsilon F_{31}^{} \right)
}{\epsilon^2}\nonumber\\
& & \hspace{-0.2cm} \displaystyle  - 4 {\cal J}\alpha \left(1 + \beta\right)
\cos \left(\epsilon F_{31}^{}\right) \frac{\sin \left(\epsilon
F_{31} \right)}{\epsilon}
\nonumber\\
& & \hspace{-0.2cm} \displaystyle  + 4 \frac{\cal J} {\sin
\delta}\alpha \left(1 + \beta\right) \sin\left(F^{}_* - \beta
F_{31}^{} +\delta
\right) \frac{\sin\left(\epsilon F_{31}^{}\right)}{\epsilon} \; ,
\end{eqnarray}
where $F^{}_* \equiv \Delta^{}_* L/\left(4 E\right)$ with
$\Delta^{}_* \equiv \Delta^{}_{31} + \Delta^{}_{32}$. Since the sign
of $\Delta^{}_*$ is always the same as that of $\Delta^{}_{31}$ or
$\Delta^{}_{32}$, it can serve as a discriminator of the neutrino
mass ordering in a medium-baseline reactor antineutrino oscillation
experiment \cite{LWX}. Of course, Eq. (49) is valid for the normal
neutrino mass ordering case. When the inverted mass hierarchy (i.e.,
$\Delta^{}_{31} <0$) is concerned, the corresponding result can be
easily obtained from Eq. (49) with the replacement $ \epsilon \to
-\epsilon$, leading us to an expression which is formally the same
as Eq. (49). As for an {\it antineutrino} beam, the expression of
$\widetilde{P}(\overline{\nu}^{}_\mu \to \overline{\nu}^{}_e)$ in
the normal hierarchy case can be directly read off from Eq. (49)
with the replacements ${\delta} \to -{\delta}$ and $A \to -A$. Note
that $A \to -A$ is equivalent to $\beta \to -\beta$, implying a
consequent change of $\epsilon$.
\begin{figure}[t]
\centerline{\includegraphics[width=13.3cm]{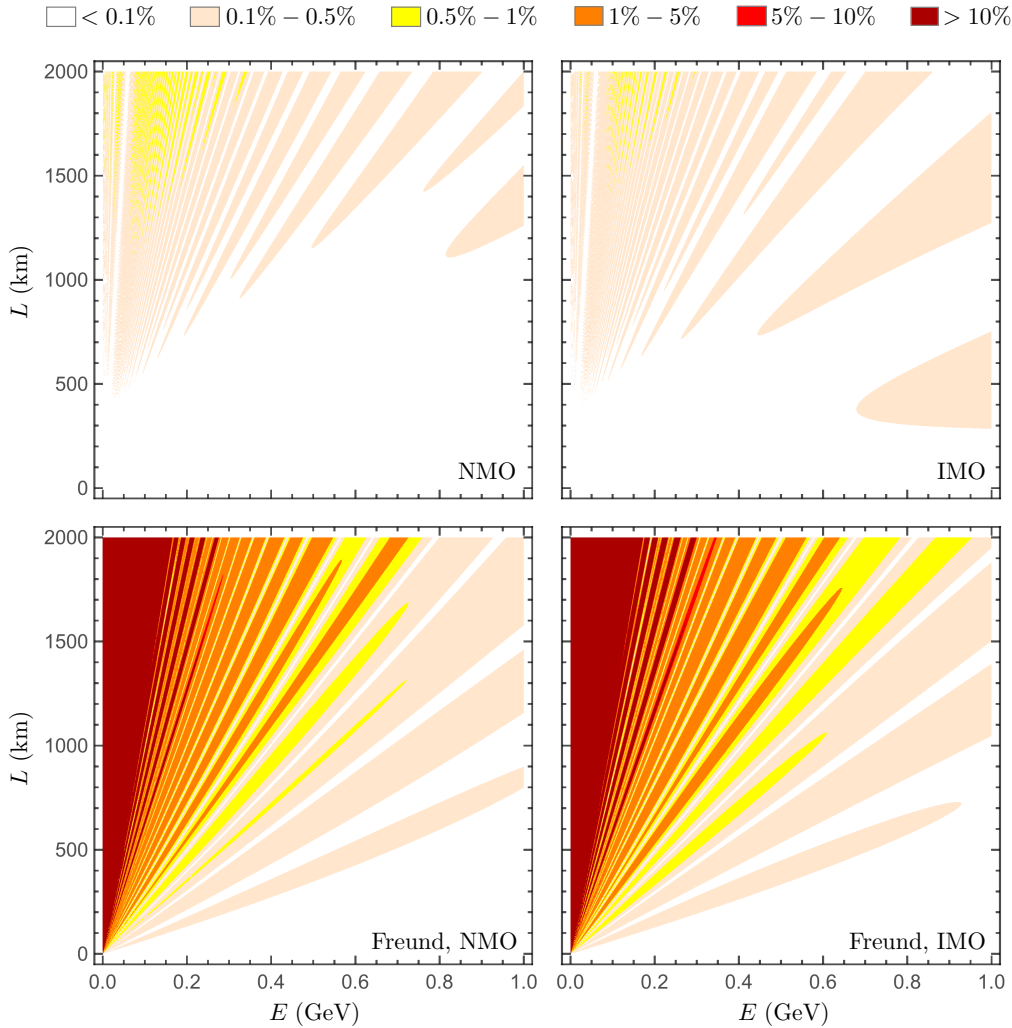}}
\caption{A comparison between the accuracies of our analytical
approximations in Eq. (49) and Freund's in Ref. \cite{Freund} by
requiring $\delta\widetilde{P}({\nu}^{}_\mu \to {\nu}^{}_e)$ defined
in Eq. (50) to be smaller than $0.1\%$, $0.1\% - 0.5\%$, $\cdots$.
Here the best-fit values of relevant oscillation parameters
\cite{GG}, together with $A \simeq 2.28 \times 10^{-4} ~{\rm eV}^2
\left(E/{\rm GeV}\right)$, have been typically input.}
\end{figure}

Different from Freund's analytical approximations for
$\widetilde{P}(\nu^{}_\mu \to \nu^{}_e)$ \cite{Freund}, which mainly
work in the $E \gtrsim 0.5$ GeV region, ours in Eq. (49) can simply
reproduce the corresponding vacuum result in the $A \to 0$ limit
(i.e., in the absence of terrestrial matter effects). Although Xu
has shown that Freund's result can be extended to cover the solar
neutrino resonance region, it is expected to be numerically less
accurate than our result. To verify this point, we illustrate the
allowed parameter space of $E$ and $L$ for a given departure of the
analytical-approximation-based numerical result of
$\widetilde{P}(\nu^{}_\mu \to \nu^{}_e)$ from the exact numerical
result in Fig. 7, in which $A \simeq 2.28 \times 10^{-4} ~{\rm eV}^2
\left(E/{\rm GeV}\right)$ is typically taken and the best-fit values
of relevant oscillation parameters \cite{GG} are input. Namely, we
require
\begin{eqnarray}
\delta \widetilde{P}(\nu^{}_\mu \to \nu^{}_e) \equiv \left|
\widetilde{P}(\nu^{}_\mu \to \nu^{}_e)^{}_{\rm exact} -
\widetilde{P}(\nu^{}_\mu \to \nu^{}_e)^{}_{\rm approximate}
\right| \lesssim 0.1\% \; , ~ 0.1\% - 0.5\% \; , \cdots ,
\end{eqnarray}
to see how small or how big the corresponding space of $E$ and $L$
is. Fig. 7 clearly shows that our analytical approximations
in Eq. (49) are numerically more accurate than Freund's in the
$E \lesssim 1$ GeV region, especially when $E$ is smaller and
smaller.

Now let us compare between the numerical results of Freund's and
ours in another way, by considering one proposed experiment (MOMENT
with $L = 150$ km \cite{MOMENT}) and two real ones (T2K with $L =
295$ km \cite{T2K} and NO$\nu$A with $L = 810$ km \cite{NOVA}).
Since the $E < 0.1$ GeV region is essentially irrelevant to these
three experiments, we have restricted ourselves to the $E\gtrsim 0.1$
GeV region in our calculations. Figs. 8 and 9 illustrate the
behaviors of $\widetilde{P}(\nu^{}_\mu \to \nu^{}_e)$ and $\delta
\widetilde{P}(\nu^{}_\mu \to \nu^{}_e)$ for the normal and inverted
neutrino mass hierarchies, respectively. We see that both Freund's
analytical approximations and ours are actually good enough to
describe the behaviors of matter-corrected $\nu^{}_\mu \to \nu^{}_e$
oscillations for the MOMENT and T2K experiments, although the
accuracy of our approximations is certainly much better. In
contrast, Freund's result is much better than ours for the NO$\nu$A
experiment, simply because the latter involves $E \gtrsim 1$ GeV. In
short, our new approximations provide an alternative analytical way
for understanding the matter-modified behaviors of $\nu^{}_\mu \to
\nu^{}_e$ and $\overline{\nu}^{}_\mu \to \overline{\nu}^{}_e$
oscillations in the $0.1 ~{\rm GeV} \lesssim E \lesssim 1 ~{\rm
GeV}$ region.
\begin{figure*}
\centerline{\includegraphics[width=15.3cm]{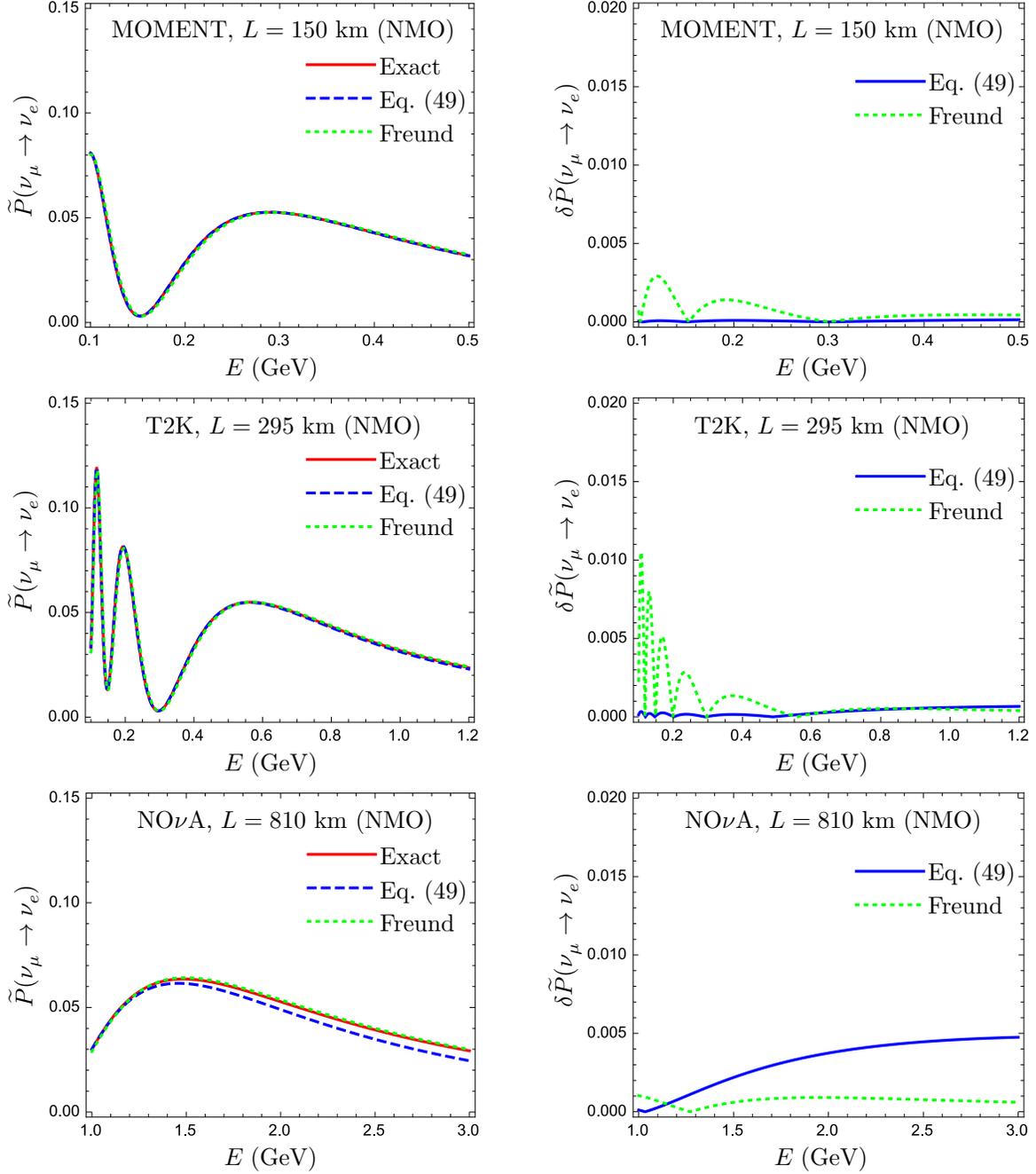}} \caption{A
comparison between the numerical accuracies of our analytical
approximations in Eq. (49) and Freund's in Ref. \cite{Freund} for
the MOMENT, T2K and NO$\nu$A experiments in the normal neutrino mass
ordering case. Here the best-fit values of relevant oscillation
parameters \cite{GG}, together with $A \simeq 2.28 \times 10^{-4}
~{\rm eV}^2 \left(E/{\rm GeV}\right)$, have been typically input.}
\end{figure*}
\begin{figure*}
\centerline{\includegraphics[width=15.3cm]{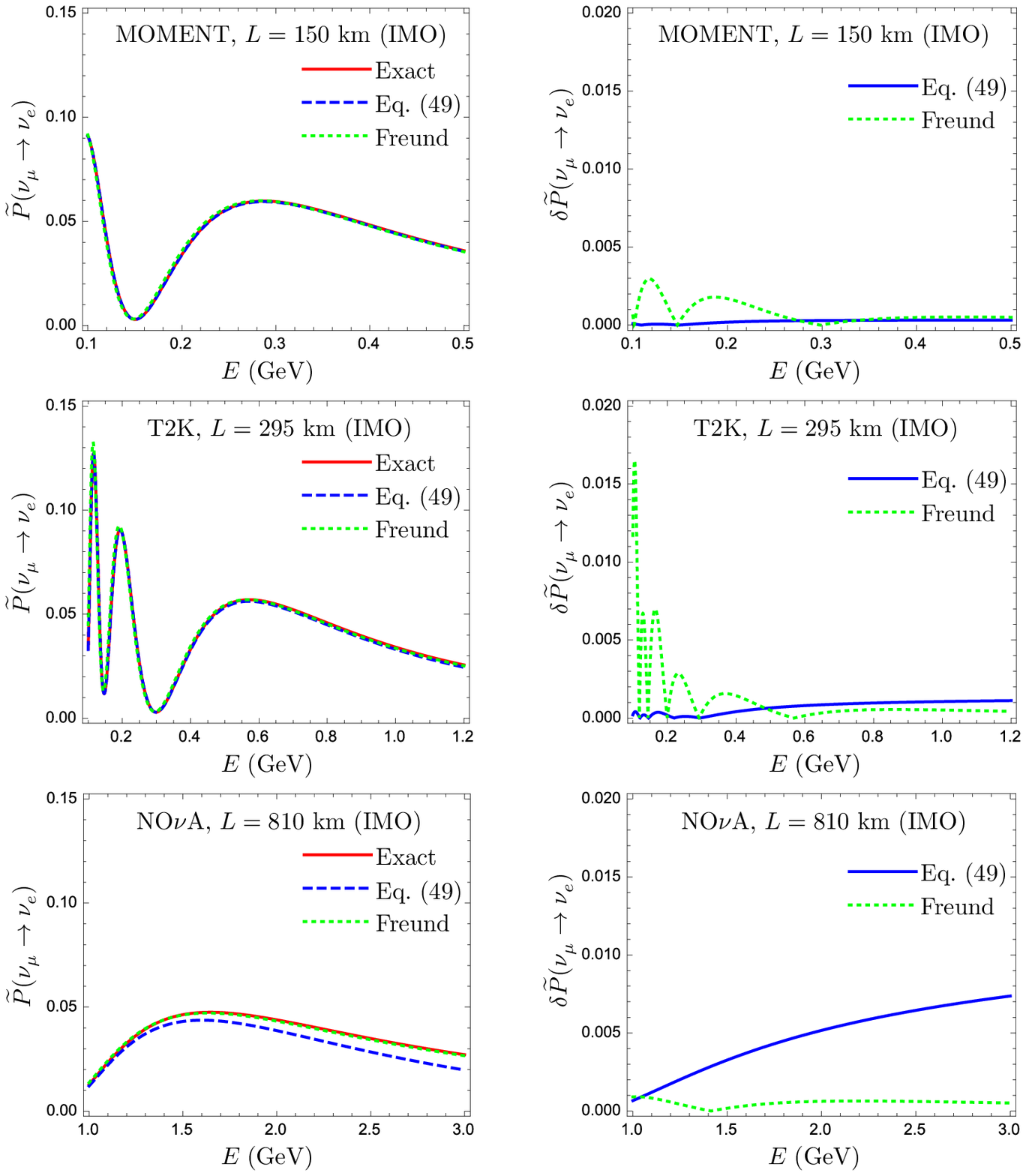}} \caption{A
comparison between the numerical accuracies of our analytical
approximations in Eq. (49) and Freund's in Ref. \cite{Freund} for
the MOMENT, T2K and NO$\nu$A experiments in the inverted neutrino
mass ordering case. Here the best-fit values of relevant oscillation
parameters \cite{GG}, together with $A \simeq 2.28 \times 10^{-4}
~{\rm eV}^2 \left(E/{\rm GeV}\right)$, have been typically input.}
\end{figure*}

In the following we focus on a low-energy medium-baseline neutrino
oscillation experiment which is capable of probing leptonic
CP-violating asymmetry
\begin{eqnarray}
\widetilde{\cal A}^{}_{\rm CP} \equiv \widetilde{\cal A}^{}_{\cal J}
+ \widetilde{\cal A}^{}_{\rm F} \equiv \widetilde{P}(\nu^{}_\mu \to
\nu^{}_e) - \widetilde{P}(\overline{\nu}^{}_\mu \to
\overline{\nu}^{}_e) \; ,
\end{eqnarray}
in which $\widetilde{\cal A}^{}_{\cal J}$ stands for the {\it
genuine} CP-violating effect governed by the nontrivial value of
Dirac phase $\delta$
\footnote{Note that $\widetilde{\cal A}^{}_{\cal J}$ as a
CP-violating asymmetry is associated with both matter ($A$) and
antimatter ($-A$), while $\widetilde{\cal A}^{}_{\rm T}$ defined in
Eq. (2) is the T-violating asymmetry and thus depends only on
matter.},
and $\widetilde{\cal A}^{}_{\rm F}$ denotes the {\it fake} asymmetry
arising from an asymmetry between terrestrial matter and antimatter.
The latter must disappear when the ``matter" parameter $A$ is
switched off. With the help of Eq. (49) and its counterpart for
$\widetilde{P}(\overline{\nu}^{}_\mu \to \overline{\nu}^{}_e)$, one
may obtain the simplified expressions of $\widetilde{\cal
A}^{}_{\cal J}$ and $\widetilde{\cal A}^{}_{\rm F}$ as
\begin{eqnarray}
\widetilde{\cal A}^{}_{\cal J} \simeq -16{\cal J} F^{}_{21} \sin^2
F^{}_{31} \simeq {\cal A}^{}_{\cal J} \equiv -16 {\cal J}
\sin F^{}_{21} \sin F^{}_{31} \sin F^{}_{32} \; ,
\end{eqnarray}
and
\begin{eqnarray}
\widetilde{\cal A}^{}_{\rm F} \hspace{-0.15cm} & \simeq &
\hspace{-0.15cm} \displaystyle 2 \beta
\left\{\sin^2 2\theta_{13}^{} \sin^2\theta_{23}^{} \left[
2 \sin^2 F_{31}^{} - (1 + \alpha) F_{31}^{} \sin
\left( 2 F_{31}^{} \right) +
\alpha \sin^2\theta_{12}^{} F_{31}^2 \cos \left( 2 F_{31}^{}
\right) \right]\right.
\nonumber\\ & & \hspace{0.55cm} \left.
- 8 \alpha \mathcal{J} \cot \delta F_{31}^2 \cos^2 F_{31}^{}
\right\} \; ,
\end{eqnarray}
if $\sin\left(\epsilon F^{}_{31}\right) \simeq \epsilon F^{}_{31}$
holds as a reasonable approximation. In this case it becomes
transparent that the fake CP-violating asymmetry $\widetilde{\cal
A}^{}_{\rm F}$ is proportional to the matter parameter $A$, while
the genuine CP-violating asymmetry $\widetilde{\cal A}^{}_{\cal J}$
in matter is essentially equal to its counterpart in vacuum. In
fact, the result in Eq. (52) is well known
\cite{Ohlsson,Minakata,Xing13}
\footnote{For example, it has been shown that the equality
$\widetilde{\cal J} \sin\widetilde{F}^{}_{21}
\sin\widetilde{F}^{}_{31} \sin\widetilde{F}^{}_{32} \simeq {\cal J}
\sin F^{}_{21} \sin F^{}_{31} \sin F^{}_{32}$ holds to a good degree
of accuracy provided the neutrino beam energy $E$ and the baseline
length $L$ satisfy the condition $10^{-7} \left(L/{\rm km}\right)^2
\left( {\rm GeV}/E\right) \ll 1$ \cite{Xing13}.},
but the one in Eq. (53) is new and instructive.
\begin{figure*}[ht]
\vspace{0.4cm} \centerline{\includegraphics[width=15.3cm]{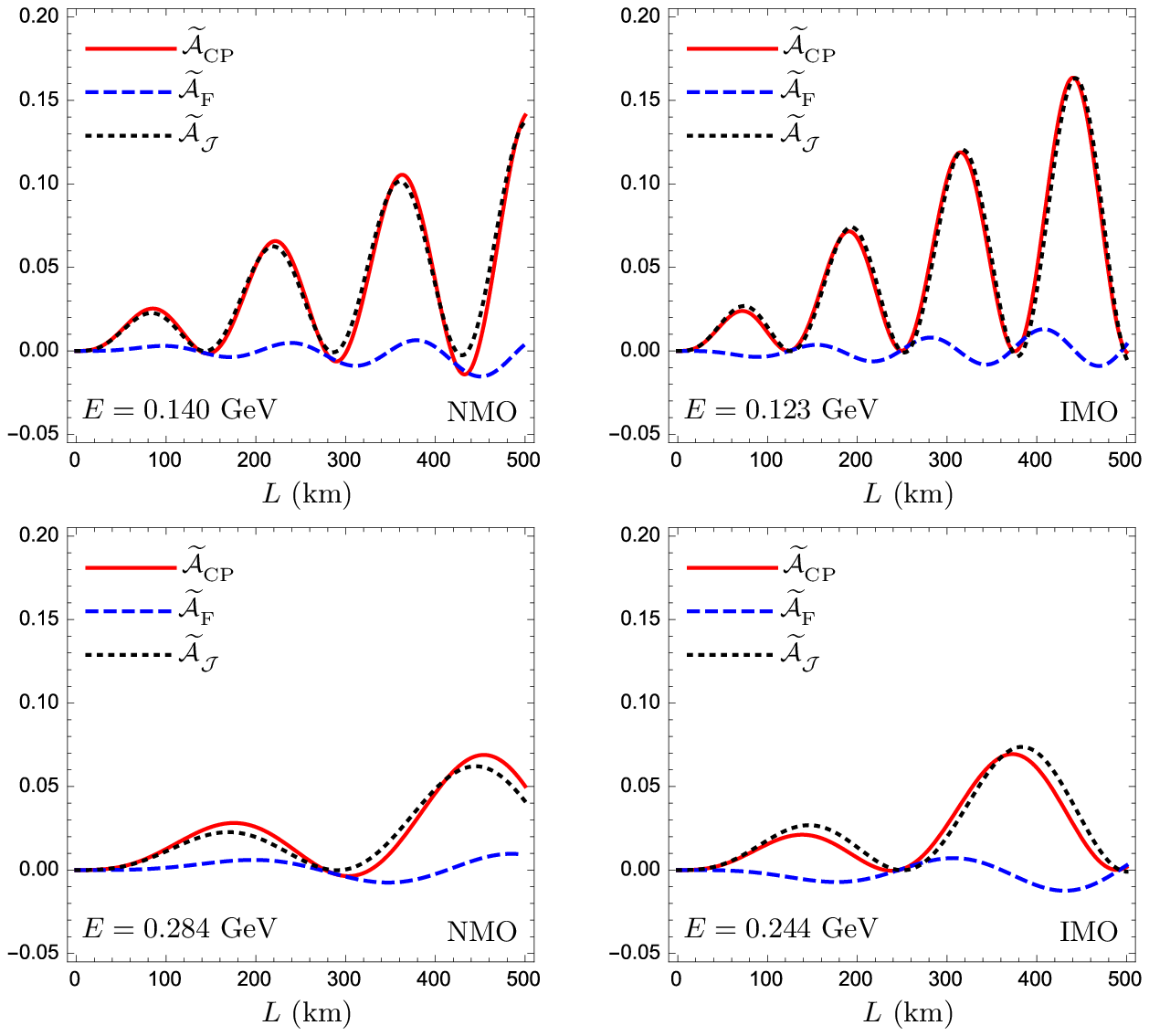}}
\caption{ The CP-violating asymmetry $\widetilde{\cal A}^{}_{\rm
CP}$ and its genuine ($\widetilde{\cal A}^{}_{\cal J}$) and fake
($\widetilde{\cal A}^{}_{\rm F}$) components for $\nu^{}_\mu \to
\nu^{}_e$ and $\overline{\nu}^{}_\mu \to \overline{\nu}^{}_e$
oscillations in matter, where the benchmark beam energies $E =
0.140$ GeV (or $0.123$ GeV) and $E = 0.284$ GeV (or $0.244$ GeV) are
taken for the normal (inverted) mass ordering, and the best-fit
values of neutrino oscillation parameters \cite{GG} have been
input.}
\end{figure*}

To illustrate the above observations in a numerical way, let us take
two benchmark values of the neutrino beam energy $E$ and plot the
asymmetries $\widetilde{\cal A}^{}_{\cal J}$, $\widetilde{\cal
A}^{}_{\rm F}$ and $\widetilde{\cal A}^{}_{\rm CP} = \widetilde{\cal
A}^{}_{\cal J} + \widetilde{\cal A}^{}_{\rm F}$ as functions of the
baseline length $L$ in Fig. 10, where the best-fit values of six
neutrino oscillation parameters have been input. These two benchmark
beam energies are just $E^{}_*$ and $E^{}_0 \simeq 2 E^{}_*$,
corresponding to the $\widetilde{\cal J}^{}_*/{\cal J}$ peak and the
nontrivial $\widetilde{\cal J}/{\cal J} =1$ point as pointed out in
section 2. One can see that $\widetilde{\cal A}^{}_{\rm CP} \simeq
\widetilde{\cal A}^{}_{\cal J}$ is an acceptable approximation in
the $E \simeq E^{}_*$ case, and the deviation of $\widetilde{\cal
A}^{}_{\rm CP}$ from $\widetilde{\cal A}^{}_{\cal J}$ can be
appreciable when $L$ becomes larger simply because the
matter-induced fake asymmetry $\widetilde{\cal A}^{}_{\rm F}$
increases with $L$ as implied in Eq. (53). Although it is possible
to obtain much larger CP-violating asymmetries when the baseline
length $L$ is properly large, a price to pay for the growth of $L$
is the decrease of the neutrino flux luminosity because the latter
is proportional to $L^{-2}$ \cite{Minakata}. For this reason, we
focus on the first two peaks of $\widetilde{\cal A}^{}_{\rm CP}$ in
Fig. 10. The values of $E$, $L$, $\widetilde{\cal A}^{}_{\rm CP}$
and $\widetilde{\cal A}^{}_{\cal J}/\widetilde{\cal A}^{}_{\rm CP}$
associated with these two peaks are summarized in Table 3. Two
comments are in order.

(a) In the case of a normal neutrino mass hierarchy,
$\widetilde{\cal A}^{}_{\cal J}/\widetilde{\cal A}^{}_{\rm CP} <1$
holds on the peaks, implying that the fake CP-violating asymmetry
$\widetilde{\cal A}^{}_{\rm F}$ contributes in a positive way.
In contrast, the contribution of $\widetilde{\cal A}^{}_{\rm F}$
is negative for the inverted neutrino mass ordering, and hence
$\widetilde{\cal A}^{}_{\cal J}/\widetilde{\cal A}^{}_{\rm CP} >1$
holds in this case.

(b) Given $E \simeq E^{}_*$ for the first peak of $\widetilde{\cal
A}^{}_{\rm CP}$, the corresponding baseline length $L$ is about
$85.20$ km (or $71.90$ km) in the $\Delta^{}_{31} >0$ (or
$\Delta^{}_{31} <0$) case. When $E \simeq 2 E^{}_*$ is taken, the
value of $L$ is roughly doubled. The situation is similar for the
second peak of $\widetilde{\cal A}^{}_{\rm CP}$. Of course, a
realistic experiment should optimize both $E$ and $L$ to make
$\widetilde{\cal A}^{}_{\rm CP}$ easily observable.
\begin{table}[t]
\centering \caption{The benchmark values of $E$, $L$,
$\widetilde{\cal A}^{}_{\rm CP}$ and $\widetilde{\cal A}^{}_{\cal
J}/\widetilde{\cal A}^{}_{\rm CP}$ associated with the first and
second peaks of the CP-violating asymmetry $\widetilde{\cal
A}^{}_{\rm CP}$ shown in Fig. 10.} \vspace{0.4cm}
\begin{tabular}[b]{c|ccccc}
\hline && \multicolumn{2}{c}{Normal mass ordering} &
\multicolumn{2}{c}{Inverted mass ordering} \\
\hline&& \multicolumn{2}{c}{$E$~(GeV)} & \multicolumn{2}{c}{
$E$~(GeV)} \\
&& 0.140 & 0.284 & 0.123 & 0.244 \\ \hline
1st peak& $\begin{array}{c} L~({\rm km})\\ \widetilde{\mathcal
A}_{\rm CP}^{}
\\ \widetilde{\mathcal A}_{\mathcal J}^{}/\widetilde
{\mathcal A}_{\rm CP}^{}\end{array}$ &
$\begin{array}{c}85.20\\ 0.025\\ 0.893\end{array}$ &
$\begin{array}{c}175.5\\ 0.028\\ 0.801\end{array}$ &
$\begin{array}{c}71.90\\ 0.024\\1.121\end{array}$ &
$\begin{array}{c}138.5\\ 0.021\\ 1.256\end{array}$\\
\hline
2nd peak & $\begin{array}{c} L~({\rm km})\\ \widetilde{\mathcal
A}_{\rm CP}^{}
\\ \widetilde{\mathcal A}_{\mathcal J}^{}/\widetilde
{\mathcal A}_{\rm CP}^{}\end{array}$ &
$\begin{array}{c}221.7\\ 0.066\\ 0.948\end{array}$ &
$\begin{array}{c}454.0\\ 0.069\\ 0.892\end{array}$ &
$\begin{array}{c}190.7\\ 0.072\\1.031\end{array}$ &
$\begin{array}{c}372.9\\ 0.069\\ 1.043\end{array}$\\
\hline
\end{tabular}
\end{table}

Furthermore, we plot the effective probabilities
$\widetilde{P}(\nu^{}_\mu \to \nu^{}_e)$ and
$\widetilde{P}(\overline{\nu}^{}_\mu \to \overline{\nu}^{}_e)$
changing with the baseline length $L$ in Fig. 11, where the inputs
are exactly the same as those used for plotting Fig. 10. Since these
two probabilities depend on $\pm A$ respectively, they receive
different contributions from terrestrial matter effects and thus
their peaks correspond to different values of $L$. The difference
between $\widetilde{P}(\nu^{}_\mu \to \nu^{}_e)$ and
$\widetilde{P}(\overline{\nu}^{}_\mu \to \overline{\nu}^{}_e)$ is
just the CP-violating asymmetry $\widetilde{\cal A}^{}_{\rm CP}$ as
illustrated in Fig. 10. Note that $\widetilde{\cal A}^{}_{\rm CP}$
is essentially insensitive to the neutrino mass hierarchy in the
leading-order approximation, because it is dominated by the
$\widetilde{\cal A}^{}_{\cal J}$ term which is insensitive to the
sign of $\Delta^{}_{31}$. This observation implies that a reasonable
determination of the CP-violating effect in the lepton sector (or
equivalently, the size of $\delta$) should in principle be possible
in such a low-energy medium-baseline neutrino oscillation experiment
even before the sign of $\Delta^{}_{31}$ is measured.
\begin{figure*}[ht]
\vspace{0.4cm} \centerline{\includegraphics[width=15.3cm]{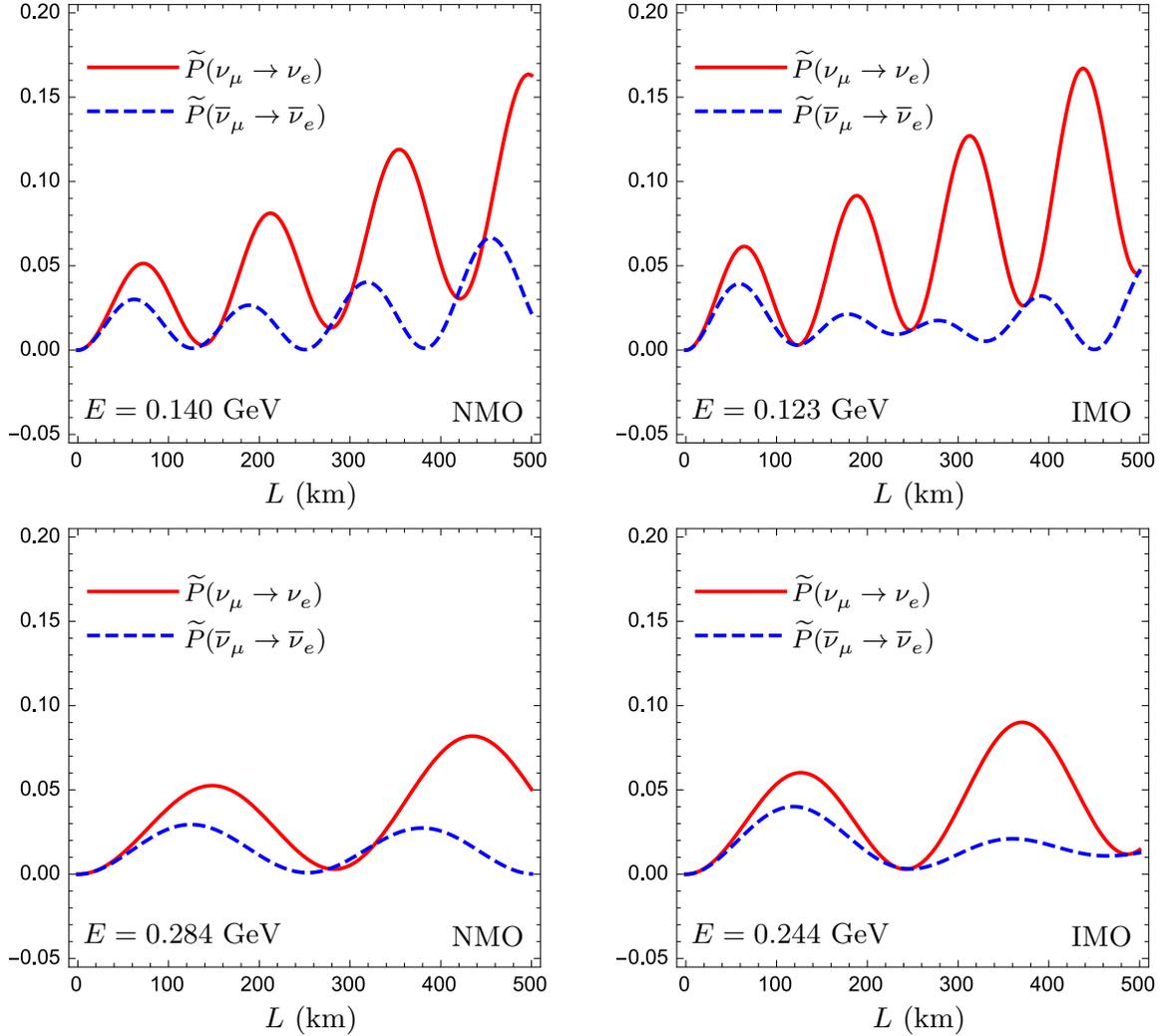}}
\caption{ The probabilities of $\nu^{}_\mu \to \nu^{}_e$ and
$\overline{\nu}^{}_\mu \to \overline{\nu}^{}_e$ oscillations in
matter, where the benchmark beam energies $E = 0.140$ GeV (or
$0.123$ GeV) and $E = 0.284$ GeV (or $0.244$ GeV) are taken for the
normal (inverted) mass ordering, and the best-fit values of neutrino
oscillation parameters \cite{GG} have been input.}
\end{figure*}

We hope that some of our results obtained in this work will be
helpful for the design of a low-energy oscillation experiment to
explore leptonic CP violation. The proposed MOMENT project
\cite{MOMENT} is just an experiment of this type. The neutrino flux
of the MOMENT is expected to peak in the $0.15 ~{\rm GeV} \lesssim E
\lesssim 0.20 ~{\rm GeV}$ region, which happens to coincide with the
$E^{}_* \lesssim E \lesssim 2 E^{}_*$ region recommended above. In
other words, this experiment is capable of probing the effects of CP
violation in $\nu^{}_\mu \to \nu^{}_e$ and $\overline{\nu}^{}_\mu
\to \overline{\nu}^{}_e$ oscillations with little matter-induced
suppression. The present studies indicate that the optimal baseline
length of the MOMENT experiment should be around $L \simeq 150$ km
\cite{MOMENT,Blennow}, which is also within the expectation shown in
Figs. 10 and 11. But the bottlenecks to the physics reach of this
experiment include how to achieve a sufficiently intense neutrino
(or antineutrino) flux and how to achieve a sufficiently high
suppression of the atmospheric neutrino background, as pointed out
and discussed in depth by Blennow {\it et al.} in Ref.
\cite{Blennow}. In this connection we plan to go into details of the
feasibility and physics potential of the MOMENT project elsewhere in
collaboration with its team members \cite{Li}.

In addition to the MOMENT facility, the ESS$\nu$SB project --- a
very intense neutrino super-beam for the measurement of leptonic CP
violation --- has recently been proposed based on the European
Spallation Source Linac \cite{ESS}. Its neutrino beam energy and
baseline length are expected to lie in the $0.2 ~{\rm GeV} \lesssim
E \lesssim 0.5 ~{\rm GeV}$ range and the $300 ~{\rm km} \lesssim L
\lesssim 600 ~{\rm km}$ range, respectively. It is obvious that the
lower-energy and shorter-baseline part of this parameter space is
consistent with our recommendation about $E$ and $L$ made above. In
fact, our analytical approximations are valid for the whole space of
$E$ and $L$ of the ESS$\nu$SB experiment, and hence they will be
very helpful to understand the numerical analysis of this
experiment's sensitivity to CP violation and matter contamination
\cite{Zhou}.

We stress that a low-energy medium-baseline neutrino oscillation
experiment can not only help probe leptonic CP violation but also
help test the other properties of lepton flavor mixing. Therefore, a
further study of this possibility is desirable \cite{Li}.

\section{Summary}

We have developed a new set of analytical approximations for the
probabilities of $\nu^{}_\mu \to \nu^{}_e$ and
$\overline{\nu}^{}_\mu \to \overline{\nu}^{}_e$ oscillations in
matter to understand the effects of leptonic CP violation in a
possible low-energy medium-baseline experiment with the beam energy
$E \lesssim 1$ GeV. Our primary motivation comes from the fact that
the previous works of this kind, such as the popular one done by
Freund \cite{Freund}, are subject to the $E \gtrsim 1$ GeV (or $E
\gtrsim 0.5$ GeV) region for a long-baseline oscillation experiment.
We have shown that our analytical approximations are numerically
more accurate than those made by Freund in the $E\lesssim 1$ GeV
region, and thus they are expected to be particularly applicable for
the MOMENT, ESS$\nu$SM and T2K experiments. The new analytical
approximations can also help us to easily understand why the
matter-corrected Jarlskog parameter $\widetilde{\cal J}$ peaks at
the resonance energy $E^{}_* \simeq 0.14$ GeV (or $0.12$ GeV) for
the normal (or inverted) neutrino mass hierarchy, and how the three
Dirac unitarity triangles are deformed due to the terrestrial matter
contamination. Finally we have affirmed that a medium-baseline
neutrino oscillation experiment with the beam energy $E$ lying in
the $E^{}_* \lesssim E \lesssim 2 E^{}_*$ range is capable of
exploring leptonic CP violation with little matter-induced
suppression.

Of course, more detailed works have to be done to combine our
analytical results with a given experiment, such as the MOMENT
project, by considering both the neutrino beam issues and the
detector issues. We plan to focus on such important but complicated
issues elsewhere in collaboration with the MOMENT team \cite{Li}.

\vspace{0.5cm}

We are indebted to Yu-Feng Li and Shun Zhou for their useful discussions
and comments. One of us (Z.Z.X.) is also grateful to Fumihiro Takayama
for his warm hospitality during the Chinese New Year at the Yukawa
Institute for Theoretical Physics of Kyoto University, where part of
this work was done. The present research is supported in part by the
National Natural Science Foundation of China under grant No. 11135009.


\newpage
\begin{thebibliography}{99}
\bibitem{PDG} K.~A.~Olive {\it et al.} [Particle Data Group Collaboration],
Review of particle physics, Chin.\ Phys.\ C {\bf 38} (2014) 090001.

\bibitem{Fogli} F.~Capozzi, G.~L.~Fogli, E.~Lisi, A.~Marrone, D.~Montanino
and A.~Palazzo, Status of three-neutrino oscillation parameters, circa 2013,
Phys.\ Rev.\ D {\bf 89} (2014) 093018 [arXiv:1312.2878 [hep-ph]].

\bibitem{Valle} D.~V.~Forero, M.~Tortola and J.~W.~F.~Valle,
Neutrino oscillations refitted, Phys.\ Rev.\ D {\bf 90} (2014) 9, 093006
[arXiv:1405.7540 [hep-ph]].

\bibitem{GG} M.~C.~Gonzalez-Garcia, M.~Maltoni and T.~Schwetz,
Updated fit to three neutrino mixing: status of leptonic CP violation,
JHEP {\bf 1411} (2014) 052 [arXiv:1409.5439 [hep-ph]].

\bibitem{T2K} K.~Abe {\it et al.} [T2K Collaboration],
Measurement of neutrino oscillation parameters from muon neutrino
disappearance with an off-axis beam, Phys.\ Rev.\ Lett.\  {\bf 111} (2013) 21,
211803 [arXiv:1308.0465 [hep-ex]].

\bibitem{T2K2} K.~Abe {\it et al.} [T2K Collaboration],
Precise measurement of the neutrino mixing parameter $\theta^{}_{23}$
from muon neutrino disappearance in an off-axis beam,
Phys.\ Rev.\ Lett.\ {\bf 112} (2014) 18, 181801 [arXiv:1403.1532 [hep-ex]].

\bibitem{NOVA} See, e.g., B. Rebel, talk given at the XIV International
Conference on Topics in Astroparticle and Underground Physics,
September 2015, Torino, Italy.

\bibitem{DYB} F.~P.~An {\it et al.} [Daya Bay Collaboration],
Spectral measurement of electron antineutrino oscillation amplitude
and frequency at Daya Bay, Phys.\ Rev.\ Lett.\ {\bf 112} (2014) 061801
[arXiv:1310.6732 [hep-ex]].

\bibitem{DYB2} F.~P.~An {\it et al.} [Daya Bay Collaboration],
New measurement of antineutrino oscillation with the full detector
configuration at Daya Bay, Phys.\ Rev.\ Lett.\ {\bf 115} (2015) 11,
111802 [arXiv:1505.03456 [hep-ex]].

\bibitem{SK} See, e.g., C. Kachulis, talk given at the EPS Conference
on High Energy Physics, July 2015, Vienna, Austria.

\bibitem{Lisi} The latest global analysis can be found in: F.~Capozzi,
E.~Lisi, A.~Marrone, D.~Montanino and A.~Palazzo, Neutrino masses
and mixings: Status of known and unknown $3\nu$ parameters,
arXiv:1601.07777 [hep-ph].

\bibitem{Wang} Y.~Wang and Z.~z.~Xing, Neutrino Masses and Flavor
Oscillations, arXiv:1504.06155 [hep-ph].

\bibitem{MOMENT} J.~Cao {\it et al.}, Muon-decay medium-baseline neutrino
beam facility, Phys.\ Rev.\ ST Accel.\ Beams {\bf 17} (2014) 090101
[arXiv:1401.8125 [physics.acc-ph]].

\bibitem{ESS} E.~Baussan {\it et al.} [ESSnuSB Collaboration],
A very intense neutrino super beam experiment for leptonic CP
violation discovery based on the European spallation source linac,
Nucl.\ Phys.\ B {\bf 885} (2014) 127 [arXiv:1309.7022 [hep-ex]].

\bibitem{Freund} M.~Freund, Analytic approximations for three neutrino
oscillation parameters and probabilities in matter, Phys.\ Rev.\ D
{\bf 64} (2001) 053003 [hep-ph/0103300].

\bibitem{Xu} X.~J.~Xu, Why is the neutrino oscillation formula expanded
in $\Delta m_{21}^{2}/ \Delta m_{31}^{2}$ still accurate near the
solar resonance in matter? JHEP {\bf 1510} (2015) 090
[arXiv:1502.02503 [hep-ph]].

\bibitem{PMNS} B.~Pontecorvo,
Mesonium and anti-mesonium, Sov.\ Phys.\ JETP {\bf 6} (1957) 429
[Zh.\ Eksp.\ Teor.\ Fiz.\  {\bf 33} (1957) 549].

\bibitem{PMNS2} Z.~Maki, M.~Nakagawa and S.~Sakata,
Remarks on the unified model of elementary particles,
Prog.\ Theor.\ Phys.\  {\bf 28} (1962) 870.

\bibitem{PMNS3} B.~Pontecorvo, Neutrino experiments and the problem of
conservation of leptonic charge, Sov.\ Phys.\ JETP {\bf 26} (1968) 984
[Zh.\ Eksp.\ Teor.\ Fiz.\  {\bf 53} (1967) 1717].

\bibitem{J} C.~Jarlskog, Commutator of the quark mass matrices in the
standard electroweak model and a measure of maximal CP violation,
Phys.\ Rev.\ Lett.\  {\bf 55} (1985) 1039.

\bibitem{MSW} L.~Wolfenstein, Neutrino oscillations in matter,
Phys.\ Rev.\ D {\bf 17} (1978) 2369.

\bibitem{MSW2} S.~P.~Mikheev and A.~Y.~Smirnov,
Resonance amplification of oscillations in matter and spectroscopy of
solar neutrinos, Sov.\ J.\ Nucl.\ Phys.\  {\bf 42} (1985) 913
[Yad.\ Fiz.\  {\bf 42} (1985) 1441].

\bibitem{Petcov} P.~I.~Krastev and S.~T.~Petcov,
Resonance amplification and T violation effects in three-neutrino
oscillations in the earth, Phys.\ Lett.\ B {\bf 205} (1988) 84.

\bibitem{Ohlsson} E.~K.~Akhmedov, P.~Huber, M.~Lindner and T.~Ohlsson,
T violation in neutrino oscillations in matter, Nucl.\ Phys.\ B {\bf 608}
(2001) 394 [hep-ph/0105029].

\bibitem{Naumov} V.~A.~Naumov, Three neutrino oscillations in matter,
CP violation and topological phases, Int.\ J.\ Mod.\ Phys.\ D {\bf 1}
(1992) 379.

\bibitem{Naumov2} P.~F.~Harrison and W.~G.~Scott,
CP and T violation in neutrino oscillations and invariance of Jarlskog's
determinant to matter effects, Phys.\ Lett.\ B {\bf 476} (2000) 349
[hep-ph/9912435].

\bibitem{Naumov3} Z.~z.~Xing, Sum rules of neutrino masses and CP violation
in the four neutrino mixing scheme, Phys.\ Rev.\ D {\bf 64} (2001) 033005
[hep-ph/0102021].

\bibitem{Minakata} H.~Minakata and H.~Nunokawa, Measuring leptonic CP
violation by low-energy neutrino oscillation experiments,
Phys.\ Lett.\ B {\bf 495} (2000) 369 [hep-ph/0004114].

\bibitem{FX00} H.~Fritzsch and Z.~z.~Xing, Mass and flavor mixing schemes
of quarks and leptons, Prog.\ Part.\ Nucl.\ Phys.\  {\bf 45} (2000) 1
[hep-ph/9912358].

\bibitem{Branco} J.~A.~Aguilar-Saavedra and G.~C.~Branco,
Unitarity triangles and geometrical description of CP violation with
Majorana neutrinos, Phys.\ Rev.\ D {\bf 62} (2000) 096009 [hep-ph/0007025].

\bibitem{Zhu} Z.~z.~Xing and J.~y.~Zhu, Leptonic unitarity triangles
and effective mass triangles of the Majorana neutrinos,
arXiv:1511.00450 [hep-ph].

\bibitem{Toshev} S.~Toshev, On T violation in matter neutrino oscillations,
Mod.\ Phys.\ Lett.\ A {\bf 6} (1991) 455.

\bibitem{Barger} V.~D.~Barger, K.~Whisnant, S.~Pakvasa and
R.~J.~N.~Phillips, Matter effects on three-neutrino oscillations,
Phys.\ Rev.\ D {\bf 22} (1980) 2718.

\bibitem{Zaglauer} H.~W.~Zaglauer and K.~H.~Schwarzer, The mixing angles
in matter for three generations of neutrinos and the MSW mechanism,
Z.\ Phys.\ C {\bf 40} (1988) 273.

\bibitem{Xing2000} Z.~z.~Xing, New formulation of matter effects on
neutrino mixing and CP violation, Phys.\ Lett.\ B {\bf 487} (2000) 327
[hep-ph/0002246].

\bibitem{Cervera} A.~Cervera, A.~Donini, M.~B.~Gavela, J.~J.~Gomez Cadenas,
P.~Hernandez, O.~Mena and S.~Rigolin, Golden measurements at a
neutrino factory, Nucl.\ Phys.\ B {\bf 579} (2000) 17
[hep-ph/0002108].

\bibitem{Li-Luo} Y.~F.~Li and S.~Luo, Neutrino oscillation probabilities
in matter with direct and indirect unitarity violation in the lepton
mixing matrix, Phys.\ Rev.\ D {\bf 93} (2016) 3, 033008
[arXiv:1508.00052 [hep-ph]].

\bibitem{matter} See, e.g., I.~Mocioiu and R.~Shrock,
Matter effects on neutrino oscillations in long baseline experiments,
Phys.\ Rev.\ D {\bf 62} (2000) 053017 [hep-ph/0002149].

\bibitem{Xing04} Z.~z.~Xing, Flavor mixing and CP violation of massive
neutrinos, Int.\ J.\ Mod.\ Phys.\ A {\bf 19} (2004) 1 [hep-ph/0307359].

\bibitem{Zhang} H.~Zhang and Z.~z.~Xing, Leptonic unitarity triangles
in matter, Eur.\ Phys.\ J.\ C {\bf 41} (2005) 143 [hep-ph/0411183].

\bibitem{Zhao} Z.~z.~Xing and Z.~h.~Zhao, A review of mu-tau flavor
symmetry in neutrino physics, arXiv:1512.04207 [hep-ph].

\bibitem{Zhang2} Z.~z.~Xing and H.~Zhang, Reconstruction of the neutrino
mixing matrix and leptonic unitarity triangles from long-baseline neutrino
oscillations, Phys.\ Lett.\ B {\bf 618} (2005) 131 [hep-ph/0503118].

\bibitem{Xing13} Z.~z.~Xing, Leptonic commutators and clean T violation in
neutrino oscillations, Phys.\ Rev.\ D {\bf 88} (2013) 017301
[arXiv:1304.7606 [hep-ph]].

\bibitem{Luo} S.~Luo, Dirac lepton angle matrix v.s. Majorana lepton
angle matrix and their renormalization group running behaviours,
Phys.\ Rev.\ D {\bf 85} (2012) 013006 [arXiv:1109.4260 [hep-ph]].

\bibitem{LWX} Y.~F.~Li, Y.~Wang and Z.~z.~Xing,
Terrestrial matter effects on reactor antineutrino oscillations at
JUNO or RENO-50: how small is small?, arXiv:1605.00900 [hep-ph].

\bibitem{Blennow} M.~Blennow, P.~Coloma and E.~Fern��ndez-Martinez,
The MOMENT to search for CP violation, arXiv:1511.02859 [hep-ph].

\bibitem{Li} Private communications with Y.~F.~Li and other members
of the MOMENT experiment.

\bibitem{Zhou} T.~Ohlsson, H.~Zhang and S.~Zhou,
Probing the leptonic Dirac CP-violating phase in neutrino
oscillation experiments, Phys.\ Rev.\ D {\bf 87} (2013) 5, 053006
[arXiv:1301.4333 [hep-ph]].
\end{thebibliography}
\end{document}